# Analysis of AWW (Anganwadi Workers) Training Content, ILA (Incremental Learning Approach) modules following CDT (Component Display Theory)

Submitted in partial fulfilment of the requirements of the degree of

**Doctor of Philosophy**

By

**Arka Majhi**

**Roll No. 184350001**

Supervisor

**Prof. Satish B. Agnihotri**

Centre for Technology Alternative for Rural Areas

Indian Institute of Technology Bombay

*November 2018*

# Seminar Approval

This is to certify that the Ph.D. Seminar report titled "**ANALYSIS OF AWW (ANGANWADI WORKERS) TRAINING CONTENT, ILA (INCREMENTAL LEARNING APPROACH) MODULES FOLLOWING CDT (COMPONENT DISPLAY THEORY)**" by **Arka Majhi** (164350001) is approved for the degree of **Doctor of Philosophy.**

| | |
|---|---|
| ____________________ | ____________________ |
| (External Examiner) | (Internal Examiner) |
| ____________________ | ____________________ |
| (Chairperson) | (Guide) |

**Date:**

**Place**: Mumbai

# Declaration

I declare that this written submission represents my ideas in my own words and where other's ideas or words have been included, I have adequately cited and referenced the original sources. I also declare that I have adhered to all principles of academic honestly and integrity and have not misrepresented or fabricated or falsified any ideas, data, facts or sources in my submission. I understand that any violation of the above will be cause of disciplinary action by the institute and evoke penal action from the sources which have thus not been properly cited or from whom proper permission has not been taken when needed.

**Date**:

**Place**: Mumbai

**Name : Arka Majhi**

**Roll No: 184350001**

# Acknowledgement

I would like to express my sincere gratitude to my guide Prof. Satish B. Agnihotri, Head of the Department, CTARA (Centre for Technology Alternatives for Rural Areas) and CPS (Centre for Policy Studies) for guiding me through each struggling stage and motivating me to keep trying and the Nutrition team (Ashish, Ayushi, Mithilesh, Rajasi and group), CTARA for supporting me throughout my project, providing me with proper guidance and resources, required to complete the seminar.

**Date**:

**Place**: Mumbai

**Name : Arka Majhi**

**Roll No: 184350001**

# Abstract

POSHAN Abhiyan, envisages, capacity building of AWWs or frontline health workers through 21 training modules of ILA (Incremental Learning Approach), modularising the net learning contents, which helps them perform their daily activities, into smaller learning topics. It focuses much on, how AWWs should perform for betterment of mother and child health from adolescence to pregnancy, during the first 1000 days of a child's development and healthy childhood. It envisions in building skilled AWWs, strengthen supervisory hierarchies, improve coordination between AWWs' (ICDS) services and health programmes to achieve common goals like increasing awareness, improving access to health and nutrition services, and reducing death and malnutrition.

For better understanding of the contents or the literature of ILA, we have conducted content analysis, by further breaking down the modules into its content types, like facts, concepts, procedure and principle. Then, we framed learning objectives to teach AWWs. We applied CDT (Component Display Theory by David Merrill) to map the contents with the desired learning objective, following the 'Specification of Objective' chart. In this way one can easily develop pedagogies from a new training literature.

The challenges in framing learning objective and pedagogies are: The AWWs do not have a (formal / scientific) nutrition and epidemiology background. Therefore, it is important to teach them through examples, familiar to them. AWWs are not evenly and structurally trained across districts. Training materials should be customized as per language, location and previous knowledge. Delayed refresher courses render them under-prepared for their jobs.

To overcome these problems, we are trying to develop an android app based on gamified learning, through which refresher training could be imparted to the AWWs. Conducting the content analysis, framing learning objectives and developing pedagogies will help conceptualizing the gamified application.

# Table of Contents

# List of Figures

# List of Tables

# Chapter 1 Introduction

## 1.1 POSHAN Abhiyan and ILA

POSHAN Abhiyaan is a programme of the Ministry of Women and Child Development (MWCD) which focuses on the first 1000 days of the child, which includes the nine months of pregnancy, six months of exclusive breastfeeding and the complementary feeding period from 6 months to 2 years to address undernutrition. It will not only help in increasing the birth weight of the children, but also help in reducing, Infant Mortality Rate (IMR) and Maternal Mortality Rate (MMR). An additional yearlong sustained intervention would consolidate the health and growth benefits of the first 1000 days. Children in the age group of 3-6 years are checked for their overall development through AWW Services.

### 1.1.1 ILA (Incremental Learning Approach)

POSHAN Abhiyaan envisages establishing a system where programme functionaries will become more effective by learning to plan and execute each task correctly and consistently through methodical, ongoing capacity building, called 'Incremental Learning Approach (ILA)'. Such a system will use opportunities in the form of existing supervisory interactions at different levels, through which practical and guided learning may be accomplished. (ICDS-WCD. ILA, 2018)

The proposed system envisages breaking down the total learning agenda into small portions of doable actions as the range of skills and tasks to be learnt is quite substantial, and since adults naturally learn by doing rather than through theory alone. The approach is to build incrementally on small amounts of learning at a time, until all skills, understanding and actions have been put into regular practice, and have been internalised by the functionaries and a supportive supervisory mechanism is put in place.

By making such a system integral to routine programme implementation, it is possible for the programme to introduce new and complex content and skills at any time and expect its effective implementation in a predictable timeline.

*Table 1.1: ILA Sequence under POSHAN Abhiyaan*

| S.No. | ILA Module name/Topic |
|---|---|
| 1 | Mapping and enumeration: Survey and listing family details– Planning and management |
| 2 | Use of home visit planner to plan and record home visits- planning and management |
| 3 | Planning and execution of community-based events at Anganwadi Center- planning and management |
| 4 | Observation of breastfeeding - breastfeeding |
| 5 | Identification and care of the weak newborn- newborn care |
| 6 | Use of home-available variety in complementary feeding- complementary feeding |
| 7 | Maternal anemia – IFA – maternal nutrition and pre- and post-natal care |
| 8 | Identification of undernutrition: Weight and height measurement- management of undernourished children |
| 9 | Complementary feeding-quantity and specificities- Complementary feeding |
| 10 | Support to exclusive breastfeeding - breastfeeding |
| 11 | Care for weak newborn- newborn care |
| 12 | Initiation of complementary feeding: Home visits in 6th and 7th months- complementary feeding |
| 13 | Identification of MAM and action required- management of Undernourished children |
| 14 | Feeding during illness – complementary feeding and follow-up on Complementary Feeding— complementary feeding |
| 15 | Supporting mothers with breastfeeding issues – breastfeeding |
| 16 | Kangaroo Mother Care for weak newborn – newborn care |
| 17 | Identification and referral of the sick newborn- newborn care |
| 18 | Hand washing and hygiene– Complementary feeding |
| 19 | Anemia in adolescent girls and children (split IL module 14) maternal nutrition and pre- and post-natal care |
| 20 | Preparation for institutional and home deliveries- maternal nutrition and pre- and post-natal care |
| 21 | Preparation during pregnancy – for newborn and family planningLevel three heading |

# Chapter 2 Content analysis

## 2.1 Need for content analysis

In order to understand the nuances of any information or literature, it needs to be classified first into its information types. All kinds of information are not perceived or internalized in the same way. Some require remembering, some require conceptual understanding, some require understand steps of a large or a critical process or procedure. In order to make a learning / instructional content, from a literature, each content types need to be dealt differently for enhancing understanding and learnability.

## 2.2 Component Display Theory (CDT)

### 2.2.1 Content classification

As per the CDT (Component Display Theory) of instructional design by M. David Merrill any learning content, can be categorized into different content types. Each of these content types corresponds to learning outcomes. Each of these learning outcomes requires different instructional strategies (Merrill, D. 1983).

- Facts are key pieces of information, such as numbers, names, etc. eg: How many IFA tablets should a pregnant mother consume in her entire pregnancy period? What is the name of the medicine, which a pregnant mother should consume for deworming?
- Concepts are symbols or objects that have similar characteristics or properties which share a name. eg: How to distinguish anemic mothers and normal mothers ?
- Procedures are a series of steps that can be used to solve the problem. eg: How to reach women before and during pregnancy ?
- Principles explain why something happens in a particular way. eg: Why are women affected more, by anemia than men ?

### 2.2.2 Performance classification

In order to learn these content types, the learner has to perform certain actions/exercises. As per the CDT, these performances can be categorized majorly into 3 types (Merrill, D. 1983).

- Remember is the performance, that requires the learner to search memory in order to reproduce or recognize some information that was previously stored
- Use is that performance that requires the learner to apply some abstractions to a specific use
- Find is that performance that requires the learner to derive or invent a new abstraction.

### 2.2.3 Content-Performance Classification

According to the CDT, all objectives or test items can be classified into one or more cells of the performance-content matrix (Merrill, D. 1983). The following chart and examples illustrate the two dimensional classification.

*Table 2.1: Content-Performance Classification*

| | | | | | |
|---|---|---|---|---|---|
| **LEVEL OF PERFORMANCE** | **FIND** | | | | |
| | **USE** | | | | |
| | **REMEMBER** | | | | |
| | | **FACT** | **CONCEPT** | **PROCEDURE** | **PRINCIPLE** |
| | | **TYPES OF CONTENT** | | | |

**Examples**

- Remember—fact

  eg: What is the name of the medicine, which a pregnant mother should consume for deworming?

Facts have no general or abstract representation, thus no Use-fact or Find-fact level in the matrix

- Remember—concept

  eg: What are the symptoms of anemia ?

- Use—concept

  eg: Is the mother as narrated in the story, suffering from anemia ?

- Find—concept

  eg: Sort the foods presented, into several categories. Match these categories into which nutrient it is rich in.

- Remember—procedure

  eg: What are the steps in filling a MCP (Mother and Child Protection) card ?

- Use—procedure

  eg: Fill the MCP card of a pregnant mother. Update it with medical details after childbirth

- Find—procedure

  eg: Devise a technique to sort anemic mothers from normal mothers

- Remember—principle

  eg: A pregnant mother is eating enough food, still she is anemic. Predict some possible hypotheses based on knowledge of nutritional value of foods.

- Use—principle

  eg: Two personas of pregnant mothers are narrated in the story. Explain with three different reasons why one could account for a better health than the other.

- Find—principle

  eg: Set up a drama to demonstrate the causes, effects and remedies of anemia.

### 2.2.4 Specifying objectives and tests for learning

Merrill created a taxonomy of learning outcomes and claimed that, there is a limited set of possible learning objectives, and those objectives of a single type, differ only in the topics that one is dealing with, but do not significantly differ in either form or substance. Hence, the task for a designer like the author is not to invent a learning objective but rather to select that objective that corresponds to the intended performance-content level (Merrill, D. 1983). This simplifies the process of specifying objectives of learning.

Table 2, Specification of Objectives in CDT by Merrill summarizes the substantive conditions, behaviour, and criteria that characterize each of the cells in performance-content matrix. The categories are listed along the left-hand column. Reading across a row, the entries in each column indicate the conditions, behaviour, and criteria that are necessary to specify an objective for that performance—content category. Each component is divided into two columns, one indicating that part of the component that is fixed or necessary for the objective to characterize the specific category and the other indicating those aspects of the component that can vary and still not affect the classification of the objective.

For example for the ‘remember-fact’ category, the fixed condition (column 3) is to present the symbol, object or event that is to be named (A), and if a set of such facts has been learned, to present the elements in random order. These two conditions are necessary for an objective to be included in the remember-fact category. The table is formed so that the reading across provides a complete statement of a given objective. Thus, reading across the ‘remember—fact’ row, an objective would be like: Given some images (column 1) of food (A) arranged in random order (column 2), the AWW will be able to recall the foods, rich in iron (B) (column 3), and sort them into two groups, (column 4) rich in iron and less/no iron with no errors and no delay (column 5) as shown by one point for each food, correctly sorted, and one point subtracted from score for each 10 seconds over 1 minute required to complete the exercise (column 6). The author was required to fill in the topic (A) and to select the specific mechanism for presenting the answer and for scoring the test, but the rest of the objective was already specified until one wants to change it for a different goal.

*Table 2.2: Specification of Objectives in CDT by Merrill (Merrill, D. 1983).*

**Given :** **of / for :** **The learner will be able to:** **by :** **with :** **as shown by**

| | CONDITIONS | | BEHAVIOR | | CRITERION | |
|---|---|---|---|---|---|---|
| | VARIABLE | FIXED | FIXED | VARIABLE | FIXED | VARIABLE |
| **USE CONCEPT** | DRAWING PICTURES DESCRIPTIONS DIAGRAMS | NEW EXAMPLES | CLASSIFY | WRITING SELECTING POINTING SORTING | SOME ERRORS SHORT DELAY | |
| **USE PROCEDURE** | WORD MATERIALS EQUIPMENT DEVICE | NAME, NEW TASK | DEMONSTRATE | MANIPULATING CALCULATING MEASURING REMOVING | SOME ERRORS TIMED OR UNTIMED | |
| **USE PRINCIPLE** | WORD DESCRIPTIONS DRAWINGS FIGURES | NAME, NEW PROBLEM | EXPLAIN OR PREDICT | PREDICTING CALCULATING DRAWING | SOME ERRORS UNTIMED | |
| **FIND CONCEPT** | DRAWINGS PICTURES DESCRIPTIONS DIAGRAMS OBJECTS | REFERENTS FROM UNSPECIFIED CATEGORIES | INVENT CATEGORIES | SORTING AND OBSERVING ATTRIBUTES SPECIFYING ATTRIBUTES | UNTIMED HIGH CORRELATION WITH OTHER USED CONCEPTS | |
| **FIND PROCEDURE** | DESCRIPTION DEMONSTRATION ILLUSTRATION SPECIFICATION | DESIRED PRODUCT OR EVENT | DERIVE STEPS | EXPERIMENT ANALYSIS TRIAL + ERROR | UNTIMED DEMONSTRATION OF UTILITY | |
| **FIND PRINCIPLE** | DESCRIPTION ILLUSTRATION OBSERVATION | EVENT | DISCOVER RELATIONSHIP | EXPERIMENT ANALYSIS OBSERVATION DEMONSTRATION | UNTIMED, APPROPIATE RESEARCH DESIGN | |
| **REMEMBER FACT** | DRAWINGS PICTURES DIAGRAMS OBJECTS | **A** IN ANY ORDER | RECALL **B** | WRITING, DRAWING, POINTING, SORTING, CIRCLING, ETC | NO ERRORS NO DELAY | 1 POINT FOR EACH CORRECT SYMBOL |
| **REMEMBER CONCEPT** | WORD SYMBOL | NAME | STATE DEFINITION | WRITING, SELECTING, CIRCLING, CHECKING, ETC | FEW ERRORS, SHORT DELAYS | 1 ERROR FOR EACH CHRACTERISTIC |
| **REMEMBER PROCEDURE** | WORD SYMBOL DIRECTIONS | NAME | STATE STEPS | DRAWING, LISTING, ORDERING ETC | FEW ERRORS, SHORT DELAYS | 1 ERROR FOR EACH STEP |
| **REMEMBER PRINCIPLE** | WORD SYMBOL | NAME | STATE RELATIONSHIP | WRITING, DRAWING, ETC | FEW ERRORS, SHORT DELAYS | 1 ERROR FOR EACH RELATIONSHIP |

# Chapter 3 Content analysis of ILA modules

## 3.1 Identification and care of a Weak Newborn baby (Module 5 & 11)

- **Identify babies likely to survive —**

  **Facts :** 30% babies born, die in a month. 10% babies are born before 8.5 months of pregnancy or birth weight less than 2 kg. 20 out of 100 of them die within 1 month. 90% babies are born after 8.5 months of pregnancy or birth weight 2 kg or more. 10 out of 90 of them die within 1 month.

  **Remember-Fact :** Ask what is the IMR of the country, state or district. How many malnourished children die in your area last year?

  **Concept :** Children born before 8.5 months of pregnancy or birth weight less than 2 kg have a higher chance of mortality than the ones born after them or have 2kg or more birthweight.

  **Concept :** How to differentiate between weak and sick babies ?

  A baby who is feeding well, but has now lost interest in breastfeeding and is inactive is a sick baby.

  **Use-Concept :** Show 5 babies and ask to identify normal, weak and sick babies.

  **Fact :** Where do we treat weak and sick babies ?

  A weak baby can be cared for at home, but a sick baby must be rushed to the hospital (SNCU or similar facility where the family can afford) to save her life.

  **Remember-Fact :** Which is the nearest SNCU near your Anganwadi area. How many weak and sick babies are there right now in your area ?

  **Procedure :** How to identify weak babies after birth ?

  - Check if the birth is before completing 8.5 months or 37 weeks. Ask the LMP and calculate EDD or use the LMP-EDD table.
  - Weigh the child and check if birthweight is less than 2kg
  - Check if the child is not able to suckle vigorously, the breast from the first day. Observe the baby breastfeeding soon after birth.

- A baby should be considered weak if any of these three points is true about the baby. Such babies need extra care in order to survive.

**Use-Procedure :** Narrate cases of 5 babies and ask AWW to identify weak babies.

**Find-Procedure :** Ask AWW to visit another AWC, call all the children around with their parents and ask her to identify the weak babies in that area. Tell het to teach the other AWW how to perform the same stepwise.

- **Calculate Date of Maturity from LMP —**

  **Facts :** A woman who starts having labor pains before Date of Maturity is likely to deliver a premature baby and should deliver in a hospital.

  **Remember-Fact :** What happens if a woman delivers before DOM?

  How many babies are delivered before DOM, in the last year in your area?

  **Concept :** Why should we record DOM along with the EDD in the home planner ?

  If the labor pain starts we can see if it is a case of premature delivery or if the DOM has already passed

  **Remember-Concept :** During labor pain how can we identify, if it is going to be a premature delivery or not ?

  **Procedure :** Find EDD, DoM from LMP (Using table)

  In the first row of the table the date of last menstrual period is given (LMP)

  In the second row the date for expected date of delivery is given (EDD)

  In the third row, the date given is for the date of maturity (DOM)

  **Use-Procedure :** Ask AWW to record an entry from a narration by a mother.

  **Find-Procedure :** Find which families to visit seeing from the register.

- **Things to observe on the date of birth —**

  **Procedure :** What to do if the baby was delivered in

  Hospital : If the family is told about premature delivery, we will ask what they suggested and if they are following them or not. Check discharge card or papers and get gestational age and birth weight and determine if the baby was born weak.

  Home : If the weighing was not done, we will weigh the baby using appropriate weighing scale and record the weight. We will check our recorded DOM and find out if it was a premature delivery or a weak baby.

How to ensure that correct birthweight is taken?
Counsel parents so that they insist hospital to measure and record birthweight.
Weight measurement can be repeated at home within the first 3 days after birth, using Salter scale or special baby weighing machine.

**Use-Procedure :** Ask AWW to perform an observation check in a home and a hospital delivery.

**Find-Procedure :** Narrate 5 cases of home and hospital deliveries and ask her to interrupt and point out if anything was wrong or missing.

- **How to help a weak baby survive? —**

  **Facts :** Premature babies require extra breastfeeding, extra warmth and extra cleanliness
  What should be done to keep the baby warm?
  She should be kept in kangaroo mother care. Room should be warm enough to make an adult sweat, particularly on cold winters, rains, or windy nights. Avoid bathing for at least one week after birth.
  **Concept :** Why should the baby be breastfed more than usual?
  Babies need to be woken up every hour to breastfeed, since they have a small tummy that cannot take much milk in one go. Early and frequent breast-feeding will provide the new-born with antibodies against infections.
  **Use-Concept :** Why babies not to breastfeed frequently in small amounts ?
  **Find-Concept :** What protects a baby under 6 months from infections ?

- **Why shouldn't the baby get bath immediately after birth? (Concept)**

  Giving a baby a bath too soon can cause hypothermia. Inside mom it was about 98.6 degree Celsius, but most babies are born in rooms that are about 70 degree Celsius. In the first few hours after birth, a baby has to use a lot of energy to keep warm. If a baby gets too cold, he or she can drop their blood sugar or have other complications.
  Babies are born covered in a white substance called vernix, which is composed of the skin cells, baby made early in development. Vernix contains proteins that prevent common bacterial infections. It is a natural antibacterial coating. Bacteria such as Group B Strep and E. coli are often transmitted to new-borns during delivery and can cause bloodstream infections, pneumonia and meningitis, and can be fatal. Vernix is nature's protection against these infections.

**Use-Concept:** Why babies, who are bathed just after birth are prone to infections?

**Find-Concept :** Why does a baby often catches cold after first birth after delivery ?

- **Why is it important to visit the weak baby's home daily for at least a week? —**

  **Facts :** What are the things we need to check on home visits of weak baby?

  Check adequate warmth (Kangaroo mother care and no bathing)

  Check adequate breastfeeding, no bottle

  Check adequate cleanliness- hand washing and cord care

  **Remember-fact:** Recall the checklist to check for on a visit to a new born child's home visit.

  **Concept :** How to understand that the baby is getting out of danger?

  Feeding vigorously at the breast is a sign that the baby is now getting out of danger.

  **Use-Concept:** A baby is born weak and stopped breastfeeding. What will help improve this danger state ?

  **Find-Concept :** A baby was born weak and stopped breastfeeding. Now she is recovering. What might have improved the condition ?

  **Facts :** What problems may happen if we stop attending their houses?

  The family may lose patience and give up or panic and do something wrong. They may start giving unnecessary home remedies or get unqualified doctors to treat the baby. We need to continue supporting and assure them as long as the baby's breastfeeding continues to improve and the baby can be woken up to feed.

## 3.2 Referral of sick newborn (Module 17)

- **Why the first month of life, is the most vulnerable period of life? (Concept)**

  New born babies are very prone to illnesses as their bodies have not yet learnt to fight infections. After spending nine months in the protection of the womb, where there are no germs, the baby suddenly comes out into a world that is not as clean.

Weak newborns, whose birth weight is low, or who are born prematurely, have a higher chance of getting infection because their bodies are weaker than healthy newborn babies.

**Use-Concept:** Why a baby is more prone to infection after birth ?

**Find-Concept :** Why a premature baby is more prone to infection after birth ?

- **What is the best protection for babies in the initial years of life? (Concept)**

  The best protection for babies after birth is the breast milk from their mothers. The mother is an adult whose body has already learnt to fight infections. Feeding breast milk will transfer this ability to fight germs in newborns.

  **Use-Concept:** A baby is born weak and infection prone. What will help fight it ?

  **Find-Concept :** A baby was born weak and stopped breastfeeding. She was exclusively breastfed for 6 months and now she had recovered well. Explain ?

- **What should the newborn babies not be fed with cup or spoon? (Concept)**

  Most of the weak newborns below 2 kg will lack stamina and energy to breastfeed in the first few days of life. Feeding them with cup or spoon will give more exposure for infection to spread if proper hygiene practices are not maintained. Weaker bodies and exposure to infections can develop very serious illnesses in newborns and that is why extra care of weaker

  **Use-Concept:** A baby is born weak and weakly breastfeeding. Should she be fed with a spoon or a cup ?

  **Find-Concept :** A baby was born weak and stopped breastfeeding. She was fed with a spoon and cup, but her health deteoriated. Explain ?

- **What should be done to prevent the infections in newborn?**
  **(Facts and Concepts)**

  Once the baby is born and has cried, immediate newborn care includes three main actions for preventing infections i.e. Breastfeeding, Warmth and Cord care, Hygiene and Clean Surroundings. While ensuring breastfeeding, warmth helps to protect newborns from infection, cord care and general hygiene and cleanliness will help in preventing infections among newborn.

- Providing timely and exclusive breastfeeding: Breast milk protects the baby from infections, and the baby who is fed early gets this protection early.
- Babies who are breastfed early are also likely to be given only breastfeeding, and caretakers are likely to avoid giving other harmful liquids thus reducing chances of infections.

  **Use-Concept:** How to avoid birth attendants from giving milk substitutes ?
- Ensuring that the baby is kept warm: This is the period during which the baby is most likely to become cold if she is not kept adequately warm and can easily catch infections. Delayed bathing, skin to skin contact and keeping the baby adequately wrapped can protect baby from getting cold.

  **Use-Concept:** Why babies, who are bathed just after birth are prone to infections?

  **Find-Concept :** Why does a baby often catches cold after first birth after delivery ?
- Ensuring clean handling of the baby and particularly of the umblical cord: Use of unsterile clamp or thread for tying the cord and unsterile blade or scissor for cutting the cord can cause infections in the baby. Similarly, applying unclean substances, like cow dung, to the cord or umbilicus can cause severe infections in the baby, which may lead to the baby's death. Nothing should be applied to the cord stump after the umblical cord has been cut. Similarly, nothing should be applied to the umbilicus or the cord until it dries up on its own.

  **Use-Concept:** Why shouldn’t we apply anything on umblical cord ?

  **Find-Concept :** Why only sterile blade is used for cutting umblical cord ?
- Washing hands is the single most effective way of preventing infection in newborns. Anyone who is touching the baby should wash hands with soap and water first.

  **Remember-fact :** Why should we wash our hands with soap before handling new born babies ?
- Essential newborn care is also the best opportunity to identify babies who need more care. Babies who are born earlier than expected date of delivery or are too small have a higher chance of getting infection and need special care to survive.

- **In the case of newborns : periods of labor, birth and immediately postbirth, are the critical times for spreading infections. A baby can get infection through : (Facts and concepts)**
    - Unsterile clamp or thread used for tying the cord.
    - Unsterile blade or knife or other instrument use for cutting the cord.
    - Application of things such as honey, ghee, turmeric, medicine, powder and specially harmful substances such as cow dung and urine, on the cord or umbilicus.
    - By wiping cord or umbilicus using unclean cloth.
    - Unclean surroundings during delivery and at home.
    - By touching the baby with unclean hands while cleaning, breastfeeding or changing clothes of baby.
    - People who are sick with cold, cough, fever, skin infection etc can also easily transfer the infection to newborns.
    - By giving food and any fluids other than breastmilk may harm the baby as it may contain germs from water or from feeding bottles or utensils.

- **How can we detect illness in newborn baby? (Concept)**

  Some of the early symptoms to observe in newborns are redness or swelling near cord, smell from the cord with pus or blood discharge. Even if they are present or not the baby might have developed severe illness. It is difficult to identify illness by just looking at the cord, infection can enter and spread rapidly and baby will look overall sick. Early recognition of the danger signs will help in identifying those babies who need urgent care and treatment.

  **Use-Concept:** How can the baby get ill even if the umblical cord looks alright ?

  **Find-Concept :** Why is it difficult to detect illness by looking at the umblical cord ?

- **What are the danger signs? (Facts and concepts)**

  Danger signs indicate serious illness and it can occur in many illnesses. Every newborn should be assessed for the presence of the following danger signs-

- Reduced Breastfeeding: If the baby was able to breastfeed properly for first few days but is showing the signs of reduced breastfeeding and disinterest, then the baby might be getting ill. It is important to keep observing such babies and refer them to hospital for treatment. babies and refer them to hospital for treatment.
- Reduced Activity: Does the baby look well? Observe his activity. A well-baby will be active and alert when awake. If the baby was active earlier but is now looking dull, drowsy, lethargic and is waking up with difficulty even after stimulation or not waking up properly then the baby has serious illness.
- Cold to touch: Check the baby's temperature; feel the temperature of baby's foot and see if it is cold to touch as compared to your own. If the baby's belly and feet are cold, the baby is too cold and may be very sick.
- Any sign of pneumonia: Observe how the baby is breathing. Unwell baby may breathe too fast, have long pauses in his breathing or will have heavy breathing. If the rate of breathing is more than 60 breaths/ minute then it is not normal.
- A sick newborn can become very sick rapidly and may not survive if there is a delay in recognition of symptoms or taking the baby to hospital. If any of these 4 danger signs is present, refer the newborn to the nearest facility where injectable antibiotics are available.
- All small babies are at risk of developing serious illnesses and need continuous assessment for these signs.

**Use-Concept:** What are the danger signs on which AWW should take special attention?

**Find-Concept :** Out of a listed symptoms, identify which are the danger signs.

- **What procedure should we follow before refering the baby; If any of the 4 danger signs are visible (Procedure)**
  - Identify hospitals beforehand where the injectible antibiotiocs are available and where there will be trained manpower available for giving injections.
  - Identify PHCs where the injectible antibiotics are available for treatment.
  - Contact ASHA and ANM immediately for providing first doze of antibiotic.

- If ASHA is available she is supposed to give first oral dose of antibiotics before sending the baby to hospital.
- If ANM is available she is supposed to give first doze of injectable antibiotics before sending the baby to hospital.
- If ASHA and ANM are not available to give the first doze in one to two hours of identifying danger signs, then take the baby directly to hospital where injectable antibiotics are available.
- Arrange for the transport and advice mother to keep the baby warm through skin – to-skin contact while traveling to hospital.
- Prepare the family that if the baby is not feeding at all, she may require admission.

**Use-Procedure :** Ask AWW to perform the procedure if she finds a child with danger signs.

**Find-Procedure :** Ask AWW to teach another AWW or health worker steps to do in case she finds a baby with danger signs.

- **Which antibiotics to be followed to treat sick newborn ? (Remember-Facts)**
  - Amoxycylin to be given orally
  - Gentamycin to be given through intramuscular injections.

## 3.3 Kangaroo Mother Care (Module 16)

- **Important steps of Kangaroo Mother Care – (Procedure)**
  - Infant should be without clothes and then kept in skin-to-skin contact with the mother.
  - Put on a cap and socks on the infant.
  - Infant should be safely wrapped to mother's body.
  - Infant should be kept in a comfortable position so that the infant is able to breathe and breastfeed with ease.
  - This technique can be adopted in all situations-lying, standing, working etc.
  - Any other member of the family can also practice KMC to take care of the weak newborn. This is necessary, so that the mother gets adequate rest.

- Monitoring: Infants under KMC need to be monitored regularly. Mother and members of the family should keep a watch on whether the neck of the infant is kept straight and s/he is breathing comfortably.
- Feed: More milk is produced, and with much ease, when the infant is closer to her mother's breasts. Considering the condition of the infant, she may be fed breast milk with the help of a small spoon.
- Excreta: An infant wets and soils her diapers several times a day. Use clean, old clothes at home to make nappies and keep changing as and when required.
- Adequate support to mother: Mother can do her daily chores while administering KMC. But she cannot do everything alone. Family members should always help her out. Any member of the family can provide KMC. Along with relieving her during KMC, the mother should be assisted in other household work, so that she gets sufficient rest and is able to provide adequate care to her infant. If not, it becomes difficult to save a weak newborn.

**Remember-Procedure :** Prepare a chart on 'steps to KMC' and put it on AWC.

**Use-Procedure :** Demonstrate using a baby doll, how should a mother perform Kangaroo Mother Care.

**Find-Procedure :** Ask the mother to perform Kangaroo Mother Care as per specified by the AWW. While AWW should interrupt, give hints and right guidance in order to rectify the mistakes done by mother.

- **How does Kangaroo Mother Care benefit the infant? (Concept)**

Kangaroo Mother Care is very beneficial in restoring a healthy and normal infancy to a weak newborn. KMC helps in following ways:

- Warmth: Body of a weak newborn rapidly loses heat, which increases the risk of her death. But Kangaroo Mother Care helps infant maintain her body temperature by being in skin-to-skin contact with the mother.
- Protection from infections: Infant is protected from infections from external sources while being under Kangaroo Mother Care.
- Ease in frequent breastfeeding: More milk is produced, and with much ease, when the infant is closer to her mother's breasts.

- Immediate identification of any inconvenience to infant: Infant is in constant vigil of the family members due to Kangaroo Mother Care, so any inconvenience to her is identified immediately.

**Remember-Concept:** What are the benefits of Kangaroo Mother Care ?

**Use-Concept :** What is the best way to ensure warmth to newborn baby ?

**Find-Concept :** A newborn child is ill. How can the mother take best care at home ?

## 3.4 Preventing illness to avert malnutrition and death (Module 18)

- **When are children more likely to fall sick? (Facts and concepts)**

First month of life: The first month of life is the most vulnerable period of life.

- New born babies are very prone to illnesses as their bodies have not yet learnt to fight infection. After spending nine months in the protection of the womb, where there are no germs, the baby suddenly comes out into a world that is not clean.
- If anything is applied to the cord or if these newborns are touched with unclean hands or are kept in unhygienic conditions, chances of infection are highly increased.
- The best protection for babies after birth is the breast milk from their mothers. The mother is an adult whose body has already learnt to fight infections. Breast milk from the mother's body contains ammunition that can fight germs.
- Weak newborns whose birth weights are low, or those who are born prematurely have higher chances of getting infections because their bodies are weaker than healthy newborn babies.
- That is why, of the children who die, majority die in the first month of their life and at the same time weak newborns are also most likely to die.
- That is also why babies, who do not get mother's milk, are most likely to get infections and die.

After 6 months until 2 years of age

- Once exclusive breastfeeding is established in the first few days after birth, the

baby remains well protected from infections for many months. That is why so fewer children fall sick when they are on exclusive breastfeeding.

- As the baby gets closer to six months, she becomes more playful and social. She learns to catch hold of things and put them in her mouth. She turns over on-to her belly and explores the floor. She is eager to make friends and happily plays with one and all. She is no longer in the protection of her mother alone. She starts picking up new germs from different people, from her toys and from the floor, and falls sick with fever or diarrhea.
- By six months of age, breast milk from the mother is no longer enough for her new requirements – she is growing bigger, playing more, and falling sick more often. She needs more food, and if sufficient complementary food is not given to her to make up for the shortfall of breast milk, she will become weak. Once she is weak, any common infection can affect her badly, and she will fall sick more often or stay sick for longer.

**Remember and use - concept :** When is the baby most likely to fall sick ? Why? What can be done at that time ?

- **What are the ways of protecting children from infections?**
  - Exclusive BF: As long as breast milk from the mother is sufficient for the child, it is perfect – untouched by hand and containing substances that help the child fight infections.
  - Avoiding bottle feeding: Serious illnesses like diarrhea and fever are common in children who are given milk in a bottle with a nipple. This is because even after cleaning with water, small amounts of milk remain in the nipple and deep in the bottle, and germs thrive on milk.
  - Immunization: Of the many diseases that children can suffer from, we now have vaccines against some of the most dangerous ones, like measles and pertussis. Immunization is done at the health sub-center and at VHSND sites
  - Vitamin A helps in reducing the severity of infections such as measles and diarrhea in children. It is available at health center or VHSND sites.
  - Food Hygiene: This includes- using fresh food to feed children, not touching the food with bare hands after cooking, using freshly washed utensils, using water pots having a tap or using a cup with a handle to get water out from a water pot without having to dip hands inside. Remember that cooking of food kills germs and cooked food is always safer than uncooked food. Even cooked food gets spoiled after few hours of cooking, re-heating and consuming this spoiled food can also cause serious illness.

- Safe Water Source: The fresh water drawn from hand pumps is safe. But if the bore well is not deep then the water will not be clean. If the source is clean then the tap water will be safe. Open defecation is the main cause of polluting the ground water with dangerous micro-organisms. The soil gets polluted with micro-organisms through the infected fecal matter of children. These micro-organisms through the polluted soil travel down deep inside the ground and pollute the ground water sources. Open defecation should be avoided to keep the water sources clean.
- Washing hands often is one of the simplest and the most effective methods to prevent infections among children - after using the toilet, before cooking, before and after feeding the child.

- **What are the steps for washing hand, or to reduce the growth of germs on hands (Procedure)**
    - WET your hands with clean, running water (warm or cold), turn off the tap and apply soap.
    - LATHER your hands by rubbing them together with the soap. Be sure to lather the back of your hands, between your fingers, and under your nails.
    - SCRUB, your hands for at least 20 seconds. Need a timer? Hum a song that lasts for 20-30 seconds, such as national anthem.
    - RINSE your hands well under clean, running water.
    - DRY your hands using a clean towel or air dry them.
    - After washing of hands it is a common practice to wipe hands with saree pallu primarily to dry them faster. This should be avoided as the saree is a potent source of infection and destroys the very purpose of hand washing. Similarly, using a hand towel repeatedly throughout the day is also not a good practice. Instead, dry hands by raising and waving them in the air, pointing upwards.

- Pointing the hands downwards immediately after washing will cause dirty water from the upper parts of the arm to run down to the clean hand and fingers and make them dirty again.

**Remember-Procedure :** Prepare a chart on 'hand washing steps' and put it on AWC.
**Use-Procedure :** Ask AWW to demonstrate the mother how to wash hands
**Find-Procedure :** Ask the mother to wash hands as per specified by the AWW. While AWW should interrupt, give hints and right guidance in order to rectify the mistakes done by mother.

## 3.5 Anaemia in children, adolescent and pregnant women (Module 7 and 19)

- **Hemoglobin, and its function —**

**Facts :** The hemoglobin (Hb) level in blood should be more than 12 for pregnant mothers. 11-12 signifies mild anemia, 10-11 signifies medium anemia and less than 9 signifies severe anemia.

**Concept :** Hb is the substance that make the blood red. Hb helps transport oxygen from our lungs to all organs through blood, by circulating in the body with every beat of heart. If oxygen is insufficient, the organ suffers, muscles get tired easily, and heart tries to pump the blood faster to send more oxygen, thus increasing the pulse rate.
**Remember—fact:** What should be the Hb level in the blood for a mother during pregnancy?
**Remember—concept:** What happens when hemoglobin in our blood is less than normal? Why do anemic patients feel lazy or irritated?

**Use—concept:** 10 MCP cards are shown with hemoglobin levels marked. Group them as per the levels of anemia.

**Find—concept:**
Human—Car—Road
Oxygen —? — Blood Stream

- **What are the different causes for anemia in women—**

**Concept:** Not consuming enough nutrients required for making hemoglobin, such as iron, folic acid, Vitamin B-12, and protein. Sometimes iron in food is not absorbed in intestines. Vitamin-C rich food helps in absorption of iron. Calcium restricts the absorption of iron. Long standing diseases such as TB, kidney stone, sickle-cell anemia or thalassemia prevent production of enough protein in Hb. When blood is lost faster than production in bone marrow like: Infestation by worms, menstruation, malaria (causes blood cell to break down), piles etc.

**Remember—concept:** What kind of food should we eat to fight anemia ? Why should we consume citric fruits like lemons during anemia ? What causes quick loss of blood? Why should one not drink milk or calcium tablet immediately after having a meal. **Use—concept:** Though I'm eating enough, still my Hb level is down; Explain why this happening is?

**Find—concept:** Plan 3-course meal for a pregnant mother ensuring iron content.

- **How long does it take to develop anemia—**

**Fact :** Iron and folic acid are stored in the body in organs like the liver and bone marrow. In a healthy person, who has been eating well, their stores can last for 6 months. **Concept :** If a person stops intake of iron, he/she will not have anemia for next six months. However, most people in our communities have just enough stores to last a few days, so they can become anemic very quickly.

**Remember—fact :** How long does it take for a person who stops eating iron rich food to develop anemia ?

**Remember—concept :** Why in India people develop anemia very quickly?

- **Symptoms to detect low Hb level, how do we diagnose for anemia—**

**Concept :** Paleness in parts of the body is due to reduction of Hb in blood. It tends to slow down neural activity and feel mentally exhausted.

**Procedure :** Check for paleness or loss of colour in palms, soles, nails, tongue and lips (sometime with black patches). Check for fast pulse or fast heart-rate. **Find—procedure:** Devise a technique to sort anemic mothers from normal mothers

- **What to do in order to prevent anemia in pregnant mothers—**

**Facts :** Ensure sufficient consumption of green leafty vegetables, cereals like bajra or ragi, beans, dry fruits (raisins, apricot). Improve absorption of iron by adding sour substances like lemon, amla, guava, orange. Prevent blood loss from diseases by taking albendazole that kill worms in our intestines, treatment for menstrual problems like piles, treating malaria early and completely.

**Concepts :** Consume food rich in protein, folic acid, Iron, Vitamin A, B-12, C (to increase absorption). Avoid consumption of tea and coffee for before and after having meals, as it reduces absorption of iron. Treat menstrual problems and diseases that destroy blood cells.

**Remember—fact :** What food should an anemic mother consume to fight anemia ? Which medicine should a pregnant mother consume for deworming ?

- **Risk and long term effect of anemia in pregnant mothers—**

**Concept:** During pregnancy, the baby inside the womb fulfills her requirement of blood through her mother. An anemic mother is not able to provide enough blood for her baby inside the womb and as a result the baby can be anemic or malnourished during birth. The mental and physical development of such children is slow. Anemic mother may become weaker due to the blood loss during delivery and may even die. **Remember—concept:** How can a baby be born anemic or malnourished ?

- **Why are women affected more, by anemia than men—**

**Principle :** Every month, women lose 30-40ml of blood due to menstruation for 30 years from adolescence to menopause around 45 years. During pregnancy, a mother need more iron to make blood for her body and of the baby.

**Concept :** Women eat last and least at home, often missing out nutritious foods that contain iron and other nutrients, such as protein rich foods, the green leafy vegetables, nuts, milk and curd.

**Remember—principle :** Why does women loose blood or require more blood even when not affected by a disease ?

**Remember—concept :** Why do most Indian women are anemic in spite of consuming enough food ?

- **What can AWWs do through ICDS program to prevent anemia in different life**

**stages—**

**Facts :** Iron tablet for every life stages

6month-5year: 1ml syrup twice a week at AWC. 5-10 year: 1 tablet twice a week. Adolescent girls: 1 tablet weekly at AWC and school respectively.

During pregnancy: Starting from second trimester, 1 tablet per day for 100 days

Post pregnancy: 1 tablet per day for 100 days

Deworming tablets once every six months for 12months– 5 years along with Vitamin A, 5-10 years, adolescents, pregnant women (1 tablet during second trimester) Advice on consumption of Iron and Vitamin C rich diet, birth spacing etc by health and ICDS frontline workers. Advice on availing incentives under Pradhan Mantri Matritva Vandana Yojana (PMMVY) to improve diet during pregnancy.

- **How to reach women before and during pregnancy—**

**How to reach adolescent girls ? What to advise them regarding preventing anemia? (Remembering—Facts)**

Prepare list of out-of-school adolescent girls, with the help of survey registers Provide 1 IFA tablet at AWC and deworming tablet once every six months. Get hemoglobin estimated through ANM. In severe anemic conditions, blood transfusion may be required.

**How to reach pregnant mothers ? What to advise them regarding preventing anemia? (Remembering—Facts)**

Ensure early registration and track them. Start supplementary food so that enough time is available for medicines. Deworming in second trimester. Advice mothers not to worry, if they experience nausea or constipation, advise them to consume 1 tablet before going to sleep and drink enough water. Advise them not to consume tea one hour before or after taking the tablet and to take something sour like Lemon, Amla, Guava.

**How to reach postnatal mothers ? What to advise them regarding preventing anemia? (Remembering—Facts)**

PNC home visits using pregnancy register. Wait for a week and initiate IFA tablets. Follow up to encourage consumption. If anemic, double the dose. Birth spacing as needed. Dietary advice

**Action plan for AWWs for IFA distribution to women post-delivery— (Procedure)** Post-natal visit on the day of birth and again in the first week and first month. Provide 30 tablets and advise them to take one tablet every day after meal. Follow up every 7 days through VHSND, home visits and THR days and check if the women is feeling better and she is taking the tablets or not.

- **What causes anemia in children— (Remember—Concepts)**

  Not initiating complementary feeding after 6 months or feeding top milk and milk from other sources. Diarrhoea and fever should be treated timely.

- **What are the effects of anemia in children—(Remember—Concepts)**

  **For Growth :** Height and weight increases in the first two years. Appropriate quantity of Hb is required for the growth of organs.

  **For Learning:** The baby in her first two years requires iron for development of her brain and body. Deficiencies would slow down her learning ability

  **For fighting against infections:** As the child grows, she often falls sick due to infections, which can be reduced by consuming variety of nutrients in the food.

## 3.6 Breastfeeding (Module 4, 10 and 15)

- **Harmful effects of giving a child anything other than breast milk during the first six months—(Remember—Concepts)**

  Giving the child anything other than breastmilk can cause diarrhoea and other such illnesses in the child. Some children die in the first few months from infections such as diarrhoea. Such deaths occur less often if the child is exclusively breastfed.

  **Use-Concept :** What causes diarrhoea in children during exclusive breastfeeding ?

- **When would, the amount of breast milk produced may get reduced during exclusive breastfeeding? What are the consequences? (Principle)**

  A child who is given liquids other than breast milk, may breastfeed less often, and this can lead to reduction in the amount of breast milk produced.

**Use and Find-principle:** Why should a child under 6 months is fed only breastmilk and no other liquids?

- **How does exclusive breastfeeding affect menstruation or repeated pregnancy? (Principle)**

  Mothers are less likely to menstruate or become pregnant again as long as they practice exclusive breastfeeding. If they start giving other liquids, they are more likely to start menstruating or become pregnant again.

  **Use and Find-principle:** In what circumstance can a mother with recent delivery get pregnant again ?

- **Can exclusive breastfeeding provide sufficient nutrition ? (Concept)**

  Breast milk has sufficient nutrients to meet the child's needs for growth and activity until the child is about six months old.

  **Use – Concept :** Does the baby under 6 months of age need to be fed with something else other than breastmilk ?

- **What to feed if the baby below 6 months is too thirsty? (Concept)**

  There is sufficient water in breast milk to meet the baby's needs. Breast milk has nine parts water out of ten. If a child is thirsty, it is better to give extra breastmilk instead of water as it is cleaner and hence safer than the water at home.

  **Use and Find - Concept :** Is there enough water in breastmilk ? Should the baby under 6 months fed water in summers ?

- **Is the mother's milk sufficient for the baby? (Concept)**

  If your baby is contented after feeding, passes urine several times a day, is gaining weight and is active and playful, she is probably getting all the milk she needs

  **Use - Concept :** How will you check if the baby is fed the right amount of milk ?

- **Should we start cow milk or powdered milk ? (Concept)**

  If the mother feels that she is not getting milk, don't start animal milk or powder milk on your own because once you start giving any milk other than your own milk, the amount of milk produced by your breasts will actually come down.

**Use - Concept :** During which practices, does the milk production comes down ?

- **What are the dangers of bottle feeding? (Concept)**

  Bottles and its nipples are very difficult to clean thoroughly; small bits of milk will remain stayed in such corners of the bottle that are extremely difficult to clean. Germs that can cause disease can easily breed in these corners and get into the baby's tummy. Babies can fall badly sick with illnesses like diarrhoea and fever. Such illnesses can kill the baby. Thousands of babies die in our country because of such illnesses from unclean feeding bottles.

  **Use - Concept :** Why is it always advised not to feed milk to the child from spoon, cups and plastic bottle/nipple? What are the possible dangers of this practice?

- **What do we do if the bottle feeding has been going on for a long time? (Procedure)**

  Stop bottle feeding immediately. Stop use of formula (powder milk) immediately. If the baby is less than six months old, switch to cup feeding of animal milk, such as buffalo or cow milk (if breastfeeding has been stopped or very much reduced since long). If the breastfeeding has been stopped all together and the child is more than six months of age, then initiate complementary feeding, and continue feeding a small amount of animal milk with cup.

  **Remember-Procedure:** How to switch from bottle feeding to healthy feeding practices?

- **How to ensure breastfeeding during birth ? (Facts)**

  This is important because families that practice giving liquids other than breast milk in the first few days are also likely to continue the practice later.

- **How to ensure breastfeeding if the baby is born at hospital ? (Facts)**

  Hospital staff can ensure early and exclusive breastfeeding before discharge from the hospital. The first home visit by the AWW/ASHA after birth should be made within a day of the mother returning home from the hospital.

- **How to ensure breastfeeding if the baby is born at home ? (Facts)**

The AWW/ASHA should try to remain present at birth, or should visit the home as soon as possible after birth.

- **How to periodically check if breastfeeding is going wrong? (Facts)**

  At every immunization visit during first six months: At the time of each vaccine, ask the mother or caregiver about exclusive breast- feeding. Anyone among ANM/AWW/ASHA should ensure that this enquiry is made.

- **When do mothers usually start feeding water or cow's milk ? (Facts)**

  Particularly around the age of 3 months: This is the time when mothers are most likely to introduce animal milk or water. During this time home visits, should be done to inquire for it. As soon as the mother comes to know that the child has fallen ill in the first six months: At such times, mothers may either stop breast feeding or start giving water or other liquids.

- **How can I increase my caring time for my baby? (Facts)**

  Take the baby with you to work. Your employer cannot prevent you from breastfeeding your baby from time to time.

- **What if I cannot take my baby to my workplace? (Facts)**

  If you cannot take your baby to your workplace, express milk in a large cup and whoever takes care of the baby at home can feed your baby with a spoon while you are away. Always make sure you breastfeed frequently at night, this will ensure that your breasts produce enough milk.

- **How to observe breastfeeding ? —**

  **Procedure :** Is the mother sitting or lying comfortably? Is the baby's body well supported for breastfeeding? Is the baby's head bent a little backward? Is the major part of areola (black area around the nipple) inside baby's mouth? Is the baby breastfeeding continuously? Is the baby sleeping while breastfeeding?

  **Concept :** If the mother is not comfortable, she will not be able to allow the baby to suckle for a long time; a healthy baby usually feeds for at least 10 minutes at a time.

  If the baby is well supported, she can feed continuously for a long time without getting tired.

If the head is bent forward, it is difficult for the baby to breathe and suckle.

With good attachment, only a small bit of the upper part of the areola will be clearly seen. Usually, a healthy, full term baby will suckle 10-15 times before pausing to swallow.

A hungry, healthy baby will not normally fall asleep before emptying out at least one breast. Even if she does fall asleep, she will wake up again and start suckling vigorously. A healthy baby can empty one breast completely and second breast at least partly during each feed.

- **How does a weak newborn breastfeed? —**

  **Concept : Is the baby well supported for feeding?** A weak newborn requires more careful support. The baby should lie supported on one hand, and the other hand should be used for positioning the breast carefully for the baby to feed easily.

  Is the major part of areola (black area around the nipple) inside baby's mouth? Is the baby breastfeeding continuously? If the baby is small, she will not be able to take the whole areola in the mouth. A small baby will suck a few times and pause to swallow and to rest

  **Is the baby sleeping while breastfeeding?** A small baby will feed slowly, take much longer to finish a feed, and will tend to fall asleep many times during a feed. She will need to be woken up repeatedly. The baby may not be able to completely empty even one breast at a time. But such babies will need to be fed more often – every hour or even at a lesser interval.

- **How to differentiate between breastfeeding by a weak newborn and a sick newborn ?**

  **Fact :** Weak babies — Some babies are born weak – either born too early or too small. They do not breastfeed as strongly as babies born at full term or having good birth weight. One can see that these babies are weak feeders right from the day on which they are born

  Sick babies — Babies can fall sick any time after birth, but usually do not fall sick in the first two days. Any baby can fall sick, whether or not she was born weak, but those who are born weak are more likely to fall sick. A baby may have been feeding well after birth, but when the baby falls sick, she loses interest in breastfeeding.

- **Concept :** A baby who is not breastfeeding vigorously from the day of birth is a baby who is born weak. Most of these babies can be cared for at home. A baby who was feeding well for the first few days, but later has lost interest in breastfeeding has probably fallen sick. Such babies can die if not immediately treated at the hospital.
  **Use-Concept :** Identify 2 babies who wis born weak and who has recently been sick.

- **Observe breastfeeding on the day of birth (Institutional Birth) —**
  **Facts :** Every month, you will have 1-2 institutional deliveries in your area. Visit the home of the newborn baby to meet the mother within a day of returning from the hospital. In most cases, the baby is weighed in the hospital and the family is counseled in case the baby is weak or if s/he requires extra care. Enquire if this has been done. In some cases, the baby may not be weighed or the family may not have been counseled on the need for special care.
  **Procedure :** Request the mother to breastfeed the baby in your presence, and quietly observe the breastfeed for 10- 15 minutes. Check if the breastfeeding is done in the right manner.
  If the baby is asleep when you visit the home, or has just been breastfed, visit again 1-2 hours later, telling the mother that you would like to observe her breastfeeding her baby

- **Observe breastfeeding on the day of birth (Home delivery) —**
  **Facts :** Sometimes, women in your area may deliver at home, and some of these deliveries may not be attended by a doctor or nurse. It becomes your responsibility to make sure that the family takes proper care of the baby at birth.
  In case of home delivery, if you have informed the family to call you as soon as the labor pains begin, then you may be able to observe all newborn care practices at birth. In any case, try and visit the family as soon as possible after delivery.
  **Procedure :** Find out whether the baby was breastfed immediately after birth, and whether anything other than the breast milk has been given to the baby.
  Request the mother to breastfeed the baby in your presence, and quietly observe the breastfeed for 10- 15 minutes. Check if the breastfeeding is done in the right manner.

If the baby is asleep when you visit the home, or has just been breastfed, visit again 1-2 hours later, telling the mother that you would like to observe her breastfeeding her baby.

**Use-Procedure :** At the day of home delivery, the AWW should observe breastfeeding by the mother. AWW should focus on critical steps for breastfeeding.

**Find-Procedure :** AWW should teach the mother steps to breastfeed during pregnancy or before delivery. After delivery AWW should come and check for critical steps in breastfeeding and rectify if done in a wrong manner.

- **How to observe breastfeeding when a baby falls sick —**

  **Facts : S**ometimes, a baby can fall sick in the first few weeks after birth. Sickness at this time can be dangerous for babies. If the sickness is serious, it is often possible to save the baby's life only by seeking treatment immediately at a hospital. If you inform families to call you home immediately if they feel the baby is not well, they will call you.

  **Procedure :** Visit the home as soon as you get such information. Request the mother to breastfeed the baby in your presence, and quietly observe the breastfeeding for 10-15 minutes. Check if the breastfeeding is done in the right manner. If the baby is asleep when you visit the home, or has just been breastfed, visit again 1-2 hours later, telling the mother that you would like to observe her breastfeeding her baby. If the baby does not wake up and take a good feed within two hours, advise the family that they must not wait any longer, and they must immediately seek care at an appropriate hospital.

  **Use-Procedure :** If a baby falls sick, the AWW should observe breastfeeding by the mother. AWW should focus on critical steps for breastfeeding.

  **Find-Procedure :** AWW should teach the mother steps to breastfeed during pregnancy or before delivery. After delivery AWW should come and check for critical steps in breastfeeding and rectify if done in a wrong manner. She should also look for symptoms to check if the baby is sick and suggest ways to treat her.

- **How to prepare for expressing breastmilk by hand —**

**Concept :** When the weak newborns or sick newborns are unable to breastfeed on their own, expressing breast milk makes it possible for the baby to have breast milk. It helps baby to meet her feeding requirement.

**Procedure :** In preparation of milk expression, following steps should be followed: The mother should first hold the baby for some time and try to breastfeed the baby. This will help in stimulation as well as in assessing if the baby is able to feed on her own.

Choose a cup of the right size and make sure it is clean: Choose a cup with a wide mouth for collecting the breast milk. Wash the cup with soap and hot water, and then air-dry the cup.

Prepare the breast for expressing the milk: Milk is produced in small glands all over the breast. Small ducts then drain the milk into the area just behind the nipples under the areola (or dark skin around the nipple), where milk collects in larger ducts. Milk collected in these large ducts can be easily pressed out of the nipple. So, the first step before pressing the milk out from the nipples is to gently get the milk from all over the breast to collect in these large ducts just behind the nipple. The mother should wash hands using hot water and soap and rinse off her breasts and nipples. She should sit comfortably, and keep the clean cup close by to collect the milk when it starts flowing. If the breast appears to be hard use a warm towel on breasts for a few minutes. This will help make the breast a little softer and less painful when pressed.

Stimulate the nipples by gently pulling or rolling the nipples with fingers. This is similar to a baby sucking at the breast and will make the milk come down on its own.

Breast Massage: Massage the breast in different ways shown in the pictures to soften the breast and release the milk. Start at the outer edges of the breast and keep pressing towards the nipples. When the breast is soft, and milk starts to flow from the nipple, it is ready to be pressed out into the cup. The mother should express the breast milk on her own as the breasts are easily hurt if another person tries. Teach a mother to express the milk herself. Touch her only to show her what to do, and be gentle

**Use-Procedure :** If the mother cannot breastfeed, the AWW should observe breastfeeding by the mother. If she still cannot breastfeed she should teach how to express breastmilk. AWW should then focus on critical steps for expressing milk by the mother.

**Find-Procedure :** AWW should teach the mother steps to breastfeed and express breastmilk during pregnancy or before delivery. After delivery AWW should come and check for critical steps in express breastfeeding and rectify if done in a wrong manner.

- **How to express breastmilk by hand —**

  **Procedure :** When the milk starts flowing out drop by drop, start expressing the milk into the clean cup:

  Place the thumb and finger opposite each other just outside the areola. (Areola is the dark coloured area around the nipple). Now press back towards the chest, and then gently squeeze the areola to express milk. Now repeat this process by changing position of thumb and fingers. Repeat rhythmically: position, push, press; position, push, press. In some mothers, at first no milk may come, but after massaging and pressing a few times, milk will start to flow. Massage first, then express. Massage again, and then express. Avoid rubbing or sliding the finger along the skin, this can be painful. The movement of the fingers should be rolling. Express milk from one breast until the flow slows and then express milk from the other breast. It can take 20 – 30 minutes to express breast milk adequately, especially in the first few days when only a little milk may be produced. The mother will require support from family members to take care of her other work and to allow her to express the milk patiently.

- **How to store, the expressed breastmilk —**

  **Procedure :** Cover the cup immediately after expressing the milk the cup immediately after expressing the milk. Store the milk in cool, hygenic and dry place. Store the milk in cool, hygenic and dry place. Stored breast milk should never be heated before feeding to the baby. Breast milk should never be heated before feeding to the baby. Breast milk should not be stored for more than 6 hours. Milk should not be stored for more than 6 hours. Don’t dip fingers in the cup, hold the cup from outside

  **Use-Procedure :** AWW should counsel mother how to store expressed milk and for how long before it gets spoiled.

- **When is the baby ready to feed expressed milk from the cup — (Fact)**

**Facts :** The baby is ready to feed from the cup if he can lick and swallow milk without coughing, choking or turning blue.

- **Using which utensils we should feed expressed milk—(Fact)**

  **Facts :** Use a small cup with a small, rounded brim or a spout to feed the baby: Choose a small size katori with a brim or cup with a spout. Make sure that the edge of cup or katori is not sharp, since that may injure the baby.

  A small baby can also be fed by using a katori and spoon. Choose a small size spoon with smooth edge.

- **How to check if the baby is awake and hungry ? (Fact)**

  **Facts :** Make sure the baby is awake before feeding the baby. Wake up the baby if asleep. Never try to pour milk into the mouth of a sleeping baby.

  Observe the baby to look for signals that the baby is hungry –

  Baby is moving his mouth and tongue. Eyes are open and she is looking around.

- **What to do with unused expressed milk ? (Fact)**

  **Facts :** Expressed milk left after feeding the baby can be reused for another feed within the next six hours. Do not use breastmilk that has been stored for more than 6 hours, throw it away.

- **Steps for assisting a weak newborn transition to breastfeeding:**

  **Procedure :** Remind the mother to practice the Kangaroo mother care. Skin to skin contact will help the baby to start breastfeeding sooner. Recognize the feeding signals and put the baby to the breasts briefly. Repeat this process everytime before feeding the expressed milk. Express one or two drops of milk directly into baby's mouth. If she chokes or coughs, then she is not ready to breastfeed yet. As the baby matures, she will stay awake longer at the breast and will keep sucking for longer periods of time. Sucking will help to develop her feeding skills. As the baby sucks directly at the breast for longer periods, she will gradually require lesser expressed milk, and will be happy with breastfeeding alone.

  Remember:

If the baby was taking expressed milk earlier but has now stopped taking it after 2 – 3 attempts , then refer the baby to the nearest health facility.

- **How to identify breast congestion?**

  **Procedure :** The breasts are swollen and painful to touch. The skin of breast may look shiny. The mother may have fever. The baby is unable to feed easily at the breast, and so is hungry and irritable.

  **Use-Procedure :** AWW should counsel mothers on how to check for breast congestion and treat it.

- **Why breasts get congested –**

  **Concepts :** This happens due to accumulation of milk in the ducts, the milk is not fully removed during breastfeeding. This can happen due to many reasons:

  The baby is not attached to the breast correctly during feeding, and so is not able to suck out the milk effectively. Breastfeeding is initiated many hours or days after birth, by which time the breasts have become overfilled with milk. The baby is a weak newborn who does not have the strength to suck strongly, or is satisfied with much less milk than what the breasts produce. The baby is being given other substances to drink which fills up her tummy and so she does not have the appetite to suck all the milk that is produced. The nipples are sore, and from the fear of pain the mother avoids feeding from that breast.

- **What to do when one or more breasts get congested ?**

  **Procedure :** Apply warm compresses to the breasts, which helps the milk flow, before emptying the breast by expressing out all the milk. Start breastfeeding on the emptied breast. Find out what caused the problem and correct it. Use paracetamol tablets to relieve the pain of mother.

- **How to identify sore nipples ?**

  **Facts :** Mother has severe nipple pain when the baby is suckling. A visible crack may appear across the tip of the nipple or around the base, or there may be a little bleeding.

- **Why mothers develop sore nipples ?**

  **Concept:** If the baby cannot take a large part of the areola in the mouth while breastfeeding, the baby tends to bite the nipple and injure it. This happens when the baby's head is held too far away from the nipple. If the baby is also given a bottle with a nipple, the baby may get confused and try to suck at the breast in the same way that it sucks at the bottle. Since the milk runs more easily from the bottle, the baby may get frustrated and bite and injure the nipple. If the skin of the nipple gets too dry, such as from the use of soap for cleaning the breast, the skin can crack and become painful.

- **What to do for painful nipples ?**

  **Procedure :** A painful nipple can make the mother avoid using that breast for feeding, and this can lead to congestion. The mother should be encouraged not to stop using the breast. If the pain is not relieved by paracetamol tablets and the mother wants to give rest to the painful breast, milk should be expressed out periodically before the breast gets congested. If the positioning of the baby's head is wrong, correct the position and continue to breastfeed. Usually, healing of the sore nipples is faster if breastfeeding is continued. Often as soon as the baby is well attached, the pain decreases. Apply some breastmilk to the sore nipple. Doing this every few hours will also help relieve pain and ensure healing. Allow the breastmilk to dry on the nipples and areola as it promotes healing. Expose the breasts to air for some time. If the breasts remain wet due to breastmilk or sweat, they can become more painful.

## 3.7 Complementary feeding (Module 6, 12 & 9)

- **Why do babies need a variety of nutrients even more than adults ? (Concept)**
  1. For Growth: Babies grow in height and weight throughout childhood, particularly in the mother's womb and in the first two years. This means the bones grow in length, muscles develop and all organs inside the body become larger. Every type of nutrient is needed for this growth: proteins, vitamins, iron, zinc, and a lot of energy (carbohydrates and fats) to help the growth.
  2. For learning: A child's brain grows to a size almost as large as an adult brain by 2 years of age. The child learns quickly, and her memory builds up as she sees, hears, and touches the world around her. For the development of the brain as well, every type of nutrient is needed, and any deficit leads to slow learning.
  3. For activity: With every month, the baby becomes more active, turning over, crawling, sitting up, standing and finally walking. She reaches out to hold objects, picks them up and releases them, becomes friendly with more people and plays with them. Without adequate activity, neither growth nor learning can be adequate. Activity needs a lot of energy, and a lot of the carbohydrate and fat is used up for this.

4. For fighting infections: As the child grows up in the first two years, she falls ill often, with fever, diarrhea, or cold and cough. For her body to learn to fight infections, she needs a variety of different nutrients in her diet. Absence of sufficient nutrition can prolong her illnesses; she can lose weight and fall seriously ill.

   **Use and Find-Concept :** Why should the family not cut down costs on food for the baby with other rising needs ? Why should babies be fed more than just enough food to survive ?

- **What are the different foods which provide different essential nutrients that a baby need to survive and remain healthy and fit ? (Fact)**

  **Facts :** Cereals,Millets : Provide a lot of carbohydrates – the energy that is required to run the body and its organs, perform all activities and to keep us warm. Tubers like potatoes also provide carbohydrates. But cereals also provide some protein in addition to carbohydrates which tubers do not.

  Pulses,Legumes, Nuts: Pulses and legumes are the main source of proteins for people who do not eat non-vegetarian food items. Proteins are for the body what bricks are for a building – the bones, muscles, organs are all made of proteins. Proteins are also vital for protecting us from infections and disease. Nuts are rich sources of many vitamins and minerals, particularly of zinc.

  Yellow-Orange Vegetables and Fruits: Yellow pumpkin, carrots, papaya, mango – where the whole fruit is yellow or orange – are good sources of Vitamin A for those who do not eat eggs and meats. Tomatoes are red but not a good source of Vitamin A. Vitamin A helps in keeping our eyes healthy, improves immunity and prevents infections like diarrhea, pneumonia, and measles from becoming very severe.

  Green Leafy Vegetables : Different kinds of green leafy vegetables are good sources of iron, Vitamin A and Vitamin B. Iron is used to make blood in the body. There are many components of vitamin B, each essential for important body functions such as making new blood or helping burn fuel.

Edible Oils All types of edible fats and oils are very rich sources of energy. Even small amounts of oils/fats can provide enough energy.

Milk, Curd : For those who do not eat meat and eggs, milk and its products provide good quality proteins. These are also a good source of Vitamin A and D, and calcium. Vitamin D and calcium are necessary for healthy bones.

Other Protein Rich Foods : Households that are non-vegetarian may continue to feed food items that are good sources of high quality protein, Vitamin A and D. Also the food should be well cooked, soft and mashed.

**Remember-fact :** AWW should organize a community event and discuss how to bring diversity in food and feed the right food for the growing child.

- **At what age can a child be given different foods? (Fact)**

  By the time the child is 6 months old, she can digest almost any food that she eats, as long as the food is soft and not spicy. The earlier the child is offered different foods, the earlier she will learn to appreciate different tastes and types of food. By the age of 1 year, the child can chew and eat most foods that are cooked at home.

- **How to start giving more variety of food to a child?**

**(Fact and procedures)**

1. Increase the number of food items added to a meal:
   - Most families will be willing to start feeding dal-rice or khichdi or roti soaked in dal to the child at the age of six months. As long as this meal is very soft and not spicy, babies will learn to eat small amounts in a few days. Such a meal should be offered 2-3 times a day.
   - Once the child has become used to eating small amounts of such simple food, the family can start adding other food items one by one.
   - Any food cooked at home can be fed to the child. Vegetables can be mashed and added to the rice or dal.
   - Oil or ghee can be added to the meal.
   - Milk or curd can be added to the meal, taking care that the meal does not become too watery.
   - The protein rich foods from animal sources can be added to the meal.

2. Offer a snack between meals
    - Fruits like papaya, mango, banana can be offered as snacks, after mashing the fruit to make it soft like pulp

As the mother feeds different types and tastes of food, she will slowly learn what the child likes and dislikes, and can accordingly decide to include it in the child's meals henceforth.

**Remember fact and procedures :** How to ensure balanced diet and food diversity for growing kids ? How to ensure that the amount of food consumption increases with the growth of the child ?

- **When should AWW approach the baby's family to talk about variety of foods: (Facts)**
    - Sixth month: We make home visits in the sixth month to prepare the family to start complementary feeding in the next month.
    - Immediately after six months: We make home visits immediately after the child completes six months to make sure that the family has started giving the child complementary foods, usually starting with rice and dal or rice and milk or curd.
    - Seventh and eighth months: We make home visits again in the seventh and eight months to see how well the child and mother has got used to complementary feeding. This is the right time to counsel the family to start feeding other foods cooked at home.

- **How should AWW demonstrate complimentary feeding to the family? (Procedure)**
    - Explain to the mother that they should join you in trying to feed their babies the food that they have brought from home.
    - Wash your own hands thoroughly with soap, and make the mothers also wash their own hands as well as those of babies.
    - Sit on the floor along with all the mothers and their babies.
    - Help each mother prepare a simple meal with rice and dal, or rice and curd, or roti and dal. Add a few drops of oil or ghee.
    - Ask the mothers to start feeding the meal to their babies.
    - After the babies have started eating, add one item from whatever other food they have brought, to the same meal – it could be a vegetable or milk or curd

or any other food that family eats. Mix a small amount, after properly mashing it up, and ask the mother to try and feed this mix. Observe how the child accepts this as the mother tries to feed her.

- Try adding another item in a similar manner to another part of the meal and offer that to the child a few times.
- Keep interacting with the mothers in between, explaining why it is important to add the new item and how to make it soft enough to be suitable for the child.
- Let the babies eat as much as they desire, just keep a mental note of what variety and amount the child has eaten by the end of the meal.
- At the end, ask the mothers to describe what their babies have eaten. Thank the mothers and tell them they may now take their babies home.

- **What is the right age to initiate complementary feeding?**

  **Fact :** On completion of 6 months.

- **What is the harm in initiating complementary feeding before 6 months?**

  **Concept :** Anything other than exclusive breast feeding can cause diarrhea.

- **What will happen if complementary feeding is not initiated immediately after the infant reaches 6 months of age?**

  **Concept :** Exclusive breast feeding cannot provide adequate nutrition to an infant after 6 months. Complementary feeding, along with breast feeding, is required to ensure adequate growth in height and weight and mental development of the infant. An infant of this age requires complementary feeding to acquire strength to play, and fight against infections too.

- **What diet should be given to the infant while initiating complementary feeding?**

  **Fact :** After 6 months, initiate complementary feeding with various foods available at home. Try to feed what the infants like to eat. Please make sure that the feed is soft, well cooked and mashed so that the infant can swallow without chewing.

- **Should we give thin soup of pulses (dal ka pani) or rice starch (chawal ka maadh) at the initial stage?**

**Concept :** No liquid should be given to the infant other than mother's milk. Liquids will fill her stomach but will not give required nutrition.

- **How should AWW and ASHA start counseling the family from the beginning of fifth or sixth month of the infant? – (Procedure)**
  1. Find out what is already being fed apart from breast milk and encourage exclusive breastfeeding until 6 months.
  2. Explain to the family that the child should be initiated complementary feeding along with breast milk after six months, and that such food should be cooked at home.
  3. Find out if there is a tradition similar to Annaprashan or Munhjhoothi in the family.
- **How to counsel the family after the infant has reached the age of 6 months– (Procedure)**
  1. Find out if complementary feeding has been started, and what is being given?
  2. Find out if the infant is being fed from a separate bowl or plate and spoon?
  4. Demonstrate how the food cooked at home can be prepared to feed the infant, and how the infant can be fed out of a separate bowl.
  5. Find out what problems are being faced by families in ensuring complementary feeding.
- **What types of food should be given in the beginning?**

  **Fact :** Start with semi-solid mixture of rice and pulses or khichdi. Mix available vegetables, curd, ghee, oil etc. which are available at home. Households that are non-vegetarian may continue to feed food items that are good sources of high quality protein, Vitamin A and D. Also the food should be well cooked, soft and mashed. Don't give thin soup of pulses (dal ka pani) or rice starch (chawal ka maadh). Don't give liquids like tea, coffee etc.

- **What we should ask or try to find out from the family of the child of 5 years? (Procedure)**
  - What else is being given to the child other than breast milk?
  - When are they planning to initiate complementary feeding?
  - If needed, we will explain about exclusive breastfeeding and complementary feeding.
  - First we need to meet the family and find out what is being fed to the infant.
  - If the infant is currently on exclusive breastfeeding and the family plans to introduce complementary feeding from the next month, then we will appreciate and tell them that they are on the right track.

- If they are not doing it right, then we will guide them.
- If the family is feeding water or animal milk to the infant, even if it is in very small quantity, we will urge them to stop it altogether.
- If the infant is being fed through a bottle with a nipple, then we will persuade the family to stop immediately as this practice can prove fatal for the infant.
- If the family has started the practice of complementary feeding, then we will advise them on the correct way to provide complementary foods.
- We will find out the date when child would complete six months, and inform the family that we will visit them again on that day.

- **If a child who is on mother's milk, is offered food, how much complimentary food he/she will eat, on an average: (Fact)**
    - By 9 months: about 200 gms or two standard bowls (katoris) in 24 hours
    - By 12 months: about 300 gms or three standard bowls (katoris) in 24 hours
    - By 18 months: about 500 gms or five standard bowls (katoris) in 24 hours
    - If a child is not breastfeeding, the child will eat much more.

- **What are the rules that a family should follow to ensure that the child is eating enough: (Fact)**

Offer food of different kinds and tastes, and learn what the child likes and dislikes

Offer food frequently to the child, depending on the age:

- 6-8 months: Offer food at least 3 times a day
- 9-12 months: Offer food at least 4 times a day
- After 12 months: Offer food at least 4 times a day, and in addition offer snacks twice a day

Do not restrict how much the child eats, keep feeding if the child is willing to eat

Let the child feed herself if she wants to, and let the child play with the food

Learn how to tell if the child is likely to be hungry, and feed the child at such times – don't go by the clock

Do not force a child to eat – it may work for a few days, but sooner or later the child figures out how to refuse.

As long as the child is not sick, she will eat well. When a child is allowed to eat as much as she wants, it is usually sufficient for the child.

These rules work well if the mother and child get used to them by the time the child is 9-10 months old.

Older children tend to become more and more obstinate and playful and then it becomes more difficult to instil a feeding habit. That is why it is important for the mother and baby to get into a good habit of eating before the child is one year old.

- **What should a mother feed a child during illness ? (Fact)**

**Case 1: During illness - Child less than 6 months**

**What will the child do ?**

If the child is very ill, s/he might reduce or completely stop intake of breast milk.

If the child is not very ill but is irritable, then the child might want to frequently have breast milk. The child will be thirsty particularly when s/he has fever.

**What should a mother do ?**

Breastfeeding will be the most acceptable food for the child. Mother should try to breastfeed more often. This will help to quench child's thirst as well as provide nourishment.

**Case 2: Child less than 6 months - After illness**

**What will the child do?**

Child will be more hungry; will demand to be breastfed more frequently.

**What should a mother do?**

Mother should feed as per the demand of the child. If the mother has been breastfeeding her child regularly, s/he will be able to produce more milk as per demand.

**Case 3: Child more than 6 months - During illness -**

**What will the child do?**

If the child has started complementary feeding, s/he is likely to reduce or stop eating. The child is likely to demand more breastfeeding.

**What should a mother do?**

Mother should breastfeed more often when the child is ill. If the child is willing to eat, mother should continue to feed all kinds of foods. If the child has lost appetite, the mother should encourage the child to eat at least small meals and should give varied, child's favourite and appetizing foods.

**Case 4: Child more than 6 months - After illness-**

**What will the child do?**

As the child recovers from the illness, his/her appetite will increase and he/she will want to eat more. If the child is not feeling hungry, it is likely that s/he has still not recovered from his/her illness.

**What should a mother do?**

Mother should feed the child as per his/her appetite. There is no need to restrict how much the child should eat. The child who is recovering from an illness requires more food, than before the illness. This is because the body tries to make up for the loss during the period of illness.

Mother should also offer the child as much variety of food as possible. This will help the child get all the nutrients s/he needs for recovery from illness and for growth.

**Case 5: If mother falls ill**

She should continue to breastfeed the child as per the demand. Most of the illnesses in the mother will not affect the child if mother continues to breast feed. If the mother is HIV infected, there is a chance that the child may also get infected. In such cases it is important to give exclusive breastfeeding for six months, since that is safer than non-exclusive breastfeeding. HIV infected mother should continue to take ART during this time.

- **How important is breast feeding for children with illness above six months (Concept)**

  Some mother continue breastfeeding children till the child is about two years of age, but some may stop breastfeeding early. When a child in this age falls sick, s/he usually finds it comfortable to breastfeed even if s/he does not want to take complementary feeding. Breastmilk is a wholesome food for the child and also a safe source of water, so it becomes very convenient for the mother as well to breastfeed when the child is sick. If a child falls sick after s/he has stopped breastfeeding, it is not possible for the mother to restart breastfeeding, and it can become difficult to manage feeding the child. Sometimes, such mothers may resort to bottle-feeding to comfort the child, and this can be dangerous. In this way, a mother who continues to breastfeed her child until 2 years has the advantage of being able to provide breastmilk when her child falls sick during this age.

- **Why is it important to continue feeding children even during illness? (Concept)**

  The major cause of stunting is poor complementary feeding and repeated illnesses. If the child falls ill repeatedly and is eating less food during and after illness then the opportunity to gain height during this period is lost.

- **Why children stop feeding? (Concept)**

  Loss of appetite during illness is one of the major causes of poor feeding.

- **How should children be fed during and after illness? (Facts)**

  Mother should continue the breastfeeding and increase the frequency of breastfeeding during and after illness.

  Mother should Increase the quantity and frequency of complementary feeding after illness and feed as per the child's appetite.

## 3.8 Community Bases Event – Annaprasan Diwas

- **What are the community based events organized at Anganwadi centers? (Facts)**
  - God-bharai (ritual for congratulating the mother on becoming pregnant)

- Annaprasan Diwas (ritual to celebrate child's introduction to cereals and foods other than breastmilk)

- **Why should we organize a community event? (Concept)**
  - To spread awareness on care of pregnant women and lactating mothers and children through social rituals, such as God-bharai and Annaprasan etc.
  - To help families, especially the husband and the mother-in-law, understand the importance of health and nutrition of children and pregnant women and lactating mothers.
  - Help create an atmosphere of cooperation to ensure that family and community members collectively take a decision to care for children and pregnant and lactating mothers.
  - To make the village community member aware of the health, nutrition and care requirements of children and pregnant and lactating mothers.

  **Find-Concept :** Suggest some other community based events which could bring about changes in other allied factors/indicators related to mother and child health.

- **How should we prepare for Annaprasan Diwas? (Procedure)**
  - List the children who are about to reach the age of 6 months.
  - Inform the day and date of Annaprasan Diwas to the family members. Invite all elders of the family along with the parents of the child.
  - Purchase and keep a bowl (katori), spoon and fruits (like banana and papaya) for the targeted children to be given as a token.
  - Motivate husband and mother-in-law to participate in the programme.
  - Request the family to bring a bowl of food prepared at home.
  - Invite village elders, other mothers, members of SHGs and Panchayat Representatives to bless the children.
  - You may organize other community events at your Anganwadi centers in the similar manner. Only the target beneficiaries and key messages will change according to the type of community based event.

  **Use-Procedure :** At the day of Annaprasan Diwas, the AWW should organize it stepwise before the celebration.

**Find-Procedure :** AWW should train other fellow AWWs how to organise such events

- **What should AWW do on Annaprasan Diwas? (Procedure)**
  - Clean the Anganwadi center. Keep a bucket of water and soap for hand-washing. Keep the utensils, required for use during the event, ready.
  - Spread mattress (dari or chatai) for family members to sit.
  - Display some items, such as daal, rice, vegetables, banana, papaya, etc., which can be prepared at home as complementary food for the baby.
  - Display the posters and charts for counselling on complementary feeding.
  - Request all mothers to wash their hand and dishes before initiation of Annaprasan.
  - Now request the mothers to mash daal-rice or milk-rice.
  - Ask mothers to initiate feeding the child in small quantity as this is the first time the child is having semi-solid food and hence she will eat less.
  - Motivate other members of the family to get involved in the feeding of child along with the parents.
  - Counsel the family members about initiation of complementary feeding, appropriate quantity and frequency of feeding in a day.
  - Counsel the non-vegetarian households that they may continue to feed the food items they consume.

**Use-Procedure :** At the day of Annaprasan Diwas, the AWW should follow these stepwise activities during the celebration.

**Find-Procedure :** AWW should find if everyone is performing as per suggested

- **What key messages should be given on Annaprasan Diwas? (Facts)**
  - It is essential to initiate complementary feeding along with breast milk after six months. During 6-8 months, feed 2-3 bowls of mashed food 2-3 times during the day. Add variety as the child gets used to complementary feeding.
  - A child gets adequate nutrition from breast milk up till the age of 6months.

- But after that, complementary food is required for continuous growth of the child. Feed a variety of well cooked, mashed and soft food to the child.
- Counsel the non-vegetarian households that they may continue to feed the food items they consume.
- Take care of cleanliness during preparation of food and feeding the child. Continue breastfeeding along with complementary feeding.
- Regularly weigh the child to monitor his/her growth and to take timely corrective action.

- **How to follow-up after Annaprasan Diwas? (Facts)**
  - Make frequent visits to the family till the mother and child are comfortable with the food and feeding practices, and, if required, counsel them on complementary feeding through demonstrations.
  - Tell the family members to be patient during feeding. Keep encouraging the child to eat. Ask the family members to inform the Anganwadi Worker in case they feel that the child is sick. Use the counseling points from the previous module.

## 3.9 Assessment of growth in children (Module 8)

- **Why is tracking the growth chart, so important? (Concept)**
  - Measuring a child's weight and height tells us if the child is underweight, stunted or wasted. These measurements are age and gender specific.
  - Measuring a child's weight and mapping it against her age (as on the growth chart), tells us if the child is underweight (or low weight for age). This, in general, tells us about the nutritional progress and growth of a child
  - Measuring a child's height and mapping it against its age (as on the growth chart) tells us if the child is stunted (or low height for age). This tells us that the child has chronic malnutrition, which is likely a result of long term suboptimal health and/or nutritional conditions.
  - Measuring a child's weight and mapping it against its height (as on the growth chart) tells us if the child is wasted (or low weight for height). This tells us that the child has acute malnutrition due to recent disease or lack of adequate food and nutrients.

- **Does the growth of a child differ in the first 5 years? (Concept)**
  The growth of a child depends on several factors (nutrition of the child, mother's nutrition, height etc.) and hence it varies from child to child. Despite these differences, there is a standard range of age appropriate height and weight for children upto 5 years, which tells us whether the child is exhibiting a healthy growth pattern. This has been established based on a large multi-country study, which includes Indian children and our growth charts are developed based on this.
- **Does ethnicity and genetics play a role in difference in growth? (Concept)**
  It has also been conclusively established that all children have same potential to grow during the initial years of life. Role of ethnicity and genetics do not have any influence on rate of growth during childhood. So irrespective of whether it is a child of short parents or a child of normal height parents, if provided proper nourishment it should approximately grow in the same manner in its first few years of life.
- **Method to measure weight of children (Procedure)**
  - Record child's age. If the child is too young to stand, place her on the type -1 weighing machine and note down the recordings.
  - If the child can sit but cannot stand, place her on type-2 weighing machine and record the weight.
  - If the child can stand alone, ask her to step onto the centre of the type-3 scale and stand still. Wait until the numbers on the display no longer change and stay fixed in the display.
  - Record the weight of the child to 0.01 kg.
  - Record this weight. For confirmation, record the weight once again. If there is a difference between the two readings, then measure the child for the third time to confirm the actual weight.

Precautions :

- When you weigh, you must ensure that the child is not moving. The child should look straight.
- The child should wear only light clothing and no socks or shoes.

- Check that the scale on weighing machine is displaying '0' before weighing the child. Do not weigh a child if the child is too sick or if he/she is physically disabled that will interfere with or give an incorrect measurement.

- **How to measure length for children less than 2 years of age or with height/length less than 85 cm (Procedure)**
  - Place your hands over the child's ears. With your arms straight, place the child's head against the base of the fixed head-end. The child should be looking straight up so that the line of sight is perpendicular to the board. Your head should be directly over the child's head. Watch the child's head to make sure it is in the correct position against the base of the fixed head-end of the infantometer.
  - When the child's position is correct, move the sliding foot piece with your right hand until it is firmly against the child's heels.
  - Take measurement to 0.1cm, no rounding should be done.
  - Record the length. For confirmation, record the length once again. If there is a difference between the two readings, then measure the child for the third time to confirm the actual length.

  Precautions

  - Place the infantometer on a hard, flat surface, such as the ground, floor or a solid table.
  - Make sure the child is lying flat and straight in the centre of the infantometer.
  - Ensure that the child is stable and is not moving.
  - Do not measure the length if the child is too sick or if he/she is physically disabled that will interfere with or give an incorrect measurement.

  **Use-Procedure :** At the day of Vajan Tyohar, the AWW should measure anthropometric data of children of all age upto 5 years, check with the WHO growth charts and decide whether she/he is malnourished and to what extent? Also advise the family what should be done?.
- **Find-Procedure :** AWW should train other fellow AWWs how to stepwise measure and check anthropometric data for malnutrition.

- **How to measure height for children above 2 years who are able to stand (Procedure)**
  - Child stands with back against the board. Head, hips and ankle should touch the panel as shown in the picture.
  - Body weight is evenly distributed on both feet/arms on the side. Ÿ Child's legs are placed together, bringing knees or ankles together. Ÿ Head is up and facing straight ahead.
  - Eyes level parallel to the ground (line of sight).
  - Bring headpiece down onto the upper most point on the head; compress the hair.
  - Take reading at the eye level.
  - Take measurement to 0.1cm. Do not do any rounding.
  - Record the height measurements. For confirmation, record the height once again. If there is a difference between the two readings, then measure the child for the third time to confirm the actual height.

  Precautions

  - Place the stadiometer on a hard, flat surface against a wall.
  - Remove child's shoes and socks. Apart from this, push aside braid/hair and remove clip, cap etc. that may interfere with the height measurement.
  - You must ensure that the child is not moving.
  - Do not measure the height if the child is too sick or if he/she is physically disabled that will interfere with or give an incorrect measurement.

- **What does weight and height tell us about the growth of a child? (Concept)**
  - Identifying Underweight children. Once you know the weight of a child then you will need to compare the weight of the child with the ideal weight of a child of that age and gender. If the weight of the child falls within the yellow or red zone in the growth chart, then the child is underweight. Please refer to the tables for easy understanding. For example: If a girl child who is 6 months of age, weighs below 5.7 kgs then she is underweight. Ideally, she should weigh between 5.8 to 10 kgs.
  - Identifying stunted children. Compare the height of the child by age and gender. If the height is below the cut-off value then the child is stunted. For Example: If height of a girl child, who is 6 months of age, is 59 cm then she is stunted and her ideal height should be 65.7 cm.

**Remember and use - concept**

- Give the following details to the participants and ask them to identify the child's nutritional status using the tables:

  Child 1: Age: 9 months; Gender: Male; Length: 70 cm: Weight: 9.2 Kg

  Child 2: Age: 12 months; Gender: Female; Length: 64 cm: Weight: 6.5 Kg

  Check if the participants are able to identify if the Child 1 is normal and Child 2 is underweight and stunted.

- **What does “weight for height” tell us about the growth of a child? (Concept)**
  - Identifying wasting in children. Compare the weight against length/height and gender. If the weight is below or in between the cut-off value then the child is wasted.

    Example: If the height of a boy is 64 cm and weight is 5 kg then there is considerable wasting

  **Remember and use – concept**

  Give the following details to the participants and ask them to identify the child's nutritional status using the tables:

    - Child 1: Age: 7 months; Gender: Male; Length: 60 cm: Weight: 6 Kg
    - Child 2: Age: 6 months; Gender: Female; Length: 64 cm: Weight: 5 Kg
    - Check if the participants are able to identify if the Child 1 is normal and Child 2 is severe acute wasted.

- **What should we do when a child falls in the red and yellow zone of growth chart?**

  If a child less than 24 months old is found to be stunted or underweight (falling in yellow or red zone of the height-for-age chart or weight-for-age), then the following actions are required: **(Facts)**

    - Inform the mother/parents of the child that their child's height is less for his age (growth is slow).
    - Tell the parents to pay greater attention to the following things

  Feed the child

    - With diverse foods available at home.

- Give small frequent meals.
- There should be a schedule for regularly feeding the child.
- Use separate bowl for feeding so that the mother could keep track of what and how much the child has been fed.

Sanitation and hygiene

- Washing hands with soap and water before preparing the food.
- Washing hands with soap and water before feeding the child.
- Washing hands with soap and water after defecation and disposal of waste.
- Use clean boiled water for drinking.
- Keep the food covered and protect from flies.
- Child should be regularly bathed and her nails should be clipped.

Tell the parents to get their child measured monthly, to monitor the child's progress. If the child is in the green zone/normal, then congratulate the parents and ask them to continue feeding the child well and ensuring sanitation and hygiene.

- **What should we do when a child has wasting or low weight for height? (Procedure)**
  - Measure the weight according to height and refer the baby to hospital who falls in the red zone of growth chart.
  - If the child is sick or severely malnourished, we will refer to the nearest hospital which has child specialist facility available.
  - If the child is hungry, and does not appear to be sick, we will refer her/him to the NRC if available.
  - If NRC is not available in the district, or the family is not willing to take the child, we can help the baby above 6 months by doing the following: **(Facts)**
    - We will advise the family to feed the child as much as she or he can eat.
    - Feed all kinds of food that is available at home.
    - We will advise adding oil or ghee to the food.
    - We will advise for ensuring variety in food.
    - Increase the frequency of feeding, 5 – 6 times in a day

- Households that are non-vegetarian can continue to consume food items that are rich in protein, Iron, Vitamin A, D and B12 .

**Remember and use – procedure**

- Identifying children for measurement
- Arrange 2 children (one boy and one girl) of 6-7 months and two children (one boy and one girl ) of 22-24 months.
- Ask the participants to follow the steps below:
  - STEP 1: note down the name, gender and age of the child.
  - STEP 2: Weigh the child following the steps explained earlier. Note down the readings.
  - STEP 3: Measure the height/length following all the precautions. Note down the readings.
  - STEP4: Repeat step1 to step 4 for the other child. Steps for measurement and interpretation of readings

**Find – concept,**

- The readings should be interpreted across different groups.

**Remember and use – concept,**

- Ask the participants to divide the children into the categories of SAM, MAM, stunted, wasted or underweight. By the end of this exercise the participants should be able to understand the different categories of malnutrition.

**Find – procedure**

- Discuss steps: How will you counsel each parent of those SAM, MAM children.

## 3.10 Identifying and preventing SAM (Severe Acute Malnutrition) (M 13)

- **How to measure stunting? Height against age: (Procedure)**

  We can identify stunting by measuring height against age this shows whether a child is gaining height as per age or not. Measure the child's height and refer to the table to see whether the child is stunted or not.

**Use-Procedure :** At the day of Vajan Tyohar, the AWW should measure anthropometric data of children of all age upto 5 years, check with the WHO growth charts and decide whether she/he is malnourished and to what extent? Also advise the family what should be done?.

**Find-Procedure :** AWW should train other fellow AWWs how to stepwise measure and check anthropometric data for malnutrition.

- **How to measure wasting: Weight against height: (Procedure)**

  Both weight and height need to be measured to assess if a child is thin or wasted. Height does not need to be measured every month because it does not change quickly. Measure the height once every 3 months during the first year of the child and then every six months in the next two years. Weight can be lost or gained quickly, so it is useful to measure it monthly until the child is three years of age Refer to the 'Weight for Height' table by WHO to see how thin the child is and how much the measured weight is less than expected for the measured height. The measured height and weight should be mapped against the growth chart for wasting. Any child between 6 months to 5 years who falls in yellow zone of the growth chart for wasting is moderately wasted, and any child who falls in red zone is severely wasted.

  **Use-Procedure :** At the day of Vajan Tyohar, the AWW should measure anthropometric data of children of all age upto 5 years, check with the WHO growth charts and decide whether she/he is malnourished and to what extent? Also advise the family what should be done?.

  **Find-Procedure :** AWW should train other fellow AWWs how to stepwise measure and check anthropometric data for malnutrition.

- **What are the antecedents of stunting ? (Concept)**

  Repeated minor illnesses. Not eating enough for a long period of time

  **Use-Concept :** How to anticipate through symptoms and previous records that the child may get stunted ?

- **What are the antecedents of wasting ? (Concept)**

  Recent major illness – TB, measles etc

All serious illnesses will cause wasting – heart disease, liver disease, cancer, etc. Social and family circumstances like broken families, chronically sick mother or father, extreme poverty and destitution, alcoholism, poor child care due to multiple parities, gender inequalities resulting in poor care of the girl child, etc.

**Use-Concept :** How to anticipate through symptoms and previous records that the child may get stunted ?

- **Why wasting is considered so dangerous ? (Concept)**

  Thinness reflects poor health and indicates that the body is not strong enough to fight infections. Wasting is an emergency that requires medical attention. In such circumstances, it is important to find out whether there is an underlying medical condition that is serious. Wasted children fall sick very easily. Wasted children are much more likely to die of illnesses such as diarrhea than normal or stunted children. The more wasted a child is, the more likely she/he is to die.

- **Do all wasted children require admission in a hospital? What are the recommended criteria for admission in a hospital for 6-59 months old children? (Facts and Concepts)**
  - Children falling in the red zone of the growth chart after measuring weight against height.
  - Appetite test to differentiate between:
    - ♦ Wasted, sick children
    - ♦ Wasted children with appetite
    - ♦ Children with no or very poor appetite, who are sick and need to be referred
  - Children suffering from fever since the last few days
  - Children suffering from diarrhea and vomiting continuously
  - Children who are unconscious or are listless
  - Children with high anemia
  - Children with pneumonia
  - Any type of severe illness
  - Children who have swelling in their hands and feet

**Use-Concept :** What criteria should be followed to decide that a wasted child should be hospitalized or not ?

**Find-Concept :** During Vajan Tyohar find some children who seems to be wasted (if any) and decide the severity of wasting by measuring anthropometric data, check against the WHO charts and following the criterion decide whether they should be hospitalized or not.

- **In case of wasted children, who are not admitted in hospital, what should be observed and advised to the parents ? (Facts)**
    - If sick, admit the child in a hospital for medical care.
    - If the child is less than 6 months old, consult pediatrician or NRC

- **How to help wasted children gain weight ? (Facts and Concepts)**
    - If the child is older than 6 months and is not sick, feed the child well and gradually increase the quantity and variety of complementary food like we have learnt earlier.
        - ♦ Give food available at home
        - ♦ Ensure the child is given a variety of food
        - ♦ Ensure density of food
        - ♦ More fat rich food
        - ♦ More frequently – 5-6 times per day
        - ♦ Continue breast feeding till 2 years
        - ♦ Ensure hygiene
    - Find out what caused the child to become so weak
    - Ensure complete immunization and supplementation of Vitamin A

- **How does stunting contributes to underweight? What to do about children who are in yellow or red zones on the growth chart for height-for-age and weight-for-age but are not wasted (children who are stunted but are not wasted and children who are underweight as per their age) ? (Concepts)**

- Children who are stunted have lost the opportunity to regain height. But these children should be cared for, because if they fall sick and start getting thin, then it can be dangerous for their life.
- For both stunted and underweight children, it is important to ensure complementary feeding, immunization, deworming and vitamin A supplementation.
- These children should be monitored regularly and should be referred if found to be losing weight or getting wasted or falling ill repeatedly.

## 3.11 Birth–preparedness for institutional and home delivery (Module 20)

- **When should a pregnant woman prepare for delivery ? (Facts)**
  Preparation should begin early in pregnancy, definitely by the me of god bharai.
- **When should AWW visit the mother and family to ensure planning for delivery? (Facts)** AWW must visit at least twice in the third trimester: once to ensure that they have started planning, and again to ensure that they have completed planning.
- **What are the things we need to figure out before institutional delivery ? (Facts)**
  Which is the right hospital? Which transportation to use? What are the phone numbers to call for the vehicle? What clothes are needed for mother and baby? Who will accompany? How much money will be required? What documents are required? Who will take care of the home when mother is in hospital?
- **What are the things we need to ensure before home delivery ? (Facts)**
  New blade and clean thread. Soap for washing hands. Clean sheet for mother to lie on during delivery (Cloth/Plastic). Clean cloth or pads for the mother. Clean well-lit place in the house to conduct delivery. Skilled birth attendant with phone numbers. Women to help.

## 3.12 Preparation during pregnancy-newborn care and family planning (M 21)

- **Where a newborn baby should be placed, before cutting the cord? (Concept)**

  Immediately after the birth of the baby and before cutting the cord, the baby should be wiped dry and then placed naked directly on the skin of the mother's chest/abdomen. The baby and mother together can then be covered with a dry cloth. This ensures that the baby gets adequate warmth from the body of the mother during the period immediately after birth. This is the period during which the baby is most vulnerable to become cold (hypothermic) if she is not kept adequately warm. This practice should be followed for all normal deliveries.

  **Use-Concept:** How to ensure warmth and hygiene while delivering the child?

- **What should be applied to the cord stump after cutting the cord? (Concept)**

  Nothing should be applied to the cord stump after the umbilical cord has been cut off. It should be left to dry. Similarly, nothing should be applied to the cord until it dries up and falls off after about a week after birth. Even after the cord falls off, nothing should be applied to umbilicus even if it appears wet. We give this strict message to mothers so that they do not apply harmful substances such as cow dung or anything else, even by mistake. Applying unclean substances like cow dung can cause severe infections in the baby, which may even lead to the baby's death.

  **Use-Concept:** Why shouldn't we apply anything on umblical cord ?

- **What is the best time to start breastfeeding after birth? (Concept)**

  Breastfeeding should be started as soon as possible after giving birth, preferably within an hour. If it is an institutional delivery, breastfeeding should be started before leaving the labor room. The earlier the baby starts suckling at the breast, the faster the breast milk is produced. Breast milk protects the baby from infections, and the baby who is breastfed early gets this protection early. Such babies also experience fewer problems with breastfeeding. They are also likely to be given only breastfeeding, and parents are likely to avoid giving other harmful liquids.

**Use-Concept:** Why shouldn't we apply anything on umblical cord ?

- **What is the earliest that a woman can get pregnant again after delivery? (Concept)**

  The earliest a woman can get pregnant again after a delivery is about six weeks or one and a half months. Becoming pregnant at six weeks is not common, but it can happen. It is difficult to say which woman will become pregnant early. It is safer for all women to use a contraceptive to avoid another pregnancy too soon.

  **Use-Concept:** Why is it important to counsel family for family planning just after delivery ?

  **Find-Concept :** A mother was rest assured that she is not going to get pregnant during her exclusive breastfeeding of first child. But still she conceived. How? What could dhe do to prevent it ?

- **What can happen if immediate new born care is neglected? (Concept)**

  Once the baby is born and has cried, immediate newborn care includes three main actions:

  - Ensuring clean handling of the baby and particularly of the umbilical cord.
  - Ensuring that the baby is kept warm. Providing immediate and exclusive breastfeeding.

  The main purpose of these actions is to prevent serious infections in the baby, which is one of the biggest causes of death among neonates. Babies who do not get adequate care immediately after birth are more likely to suffer severe infections and death. There are other benefits as well.

  - Immediate newborn care is also the best opportunity to identify babies who need more care.

- Babies who are born too soon or too small need special care to survive. The earlier such 'weak' babies are identified, the faster we can provide extra care. We have learned about this in previous modules.

**Use-Concept:** How is warmth and hygiene of the child taken care of just after the delivery?

**Find-Concept :** Immediate newborn care is helpful for weak, sick and premature babies. Explain?

- **What can happen if the contraception is neglected? (Concept)**

A woman can become pregnant any time after 6 weeks of delivery – it is difficult to say when she may get pregnant again. If the couple don't plan to use contraception to prevent another pregnancy early, they may end up having an unplanned pregnancy. An early and unwanted pregnancy can lead to a lot of problems:

- It will affect the breastfeeding and care of the previous child. The previous child may suffer from malnutrition or more frequent illnesses.
- It will affect the health of the mother who has not yet recovered from the previous pregnancy, and in turn will affect the development of the child in the womb – the child of the second pregnancy may be born too soon or too small.
- Parents may want an abortion and an improperly conducted abortion may cause complications or even death of the mother.

Why reach such a situation when unwanted pregnancy can be avoided using contraception!

**Use-Concept:** Why is it important to counsel family for family planning just after delivery ?

**Find-Concept :** A mother was rest assured that she is not going to get pregnant during her exclusive breastfeeding of first child. But still she conceived. How? What could she do to prevent it ?

How the mother and newborn child’s health will get affected by the second unplanned pregnancy ?

- **What preparation can AWW do during pregnancy to ensure immediate newborn care at birth? (Procedure)**

  The preparation for essential newborn care should start from the time the mother gets pregnant, and more vigorously in the third trimester, when she seriously starts planning for the delivery. We are aware that we should make at least two home visits during the last trimester. During the first home visit, we should enquire about how the mother intends to take care of her baby immediately after birth and:

  - Who in the family is likely to be with the mother during delivery? Who is likely to take care of the baby in the first hour when the mother is still recovering from labor? Has the family prepared for the 'five cleans'?
  - Do they have enough clothes to keep the baby warm?
  - When does she plan to start breastfeeding? Is the family planning to give the child any pre-lacteals? What was done when the previous child was born?
  - Is the family planning to apply anything to the cord after it is cut off? What did they do in the case of the previous delivery?

  After understanding what they plan to do, we can advise the family on correct practices, in case the plan includes anything which may be harmful to the baby. During the second visit, we can confirm if they are prepared to provide correct care to the baby at birth. We can also let them know that if the delivery happens at home for any reason, they should let us know quickly, so that we can be present at the time of birth and assist the family. When labor pains begin and the family leaves for delivery to the hospital, we will remind them again about immediate newborn care.

  **Remember-Procedure :** Prepare a checklist to follow for the mothers in the third trimester to ensure healthy delivery and put it on the wall at AWC.

  **Use-Procedure :** Ask AWW to discuss the steps she follows to counsel third trimester mothers during the home visits

  **Find-Procedure :** Ask the mother to recollect the steps told by the AWW. While AWW should interrupt, give hints and right guidance in order to rectify the mistakes done by the mother.

- **What can AWW do at birth to ensure immediate newborn care? (Procedure)**

In the case of **home delivery**, we will try and be present at birth to ensure that the family is actually practicing correct newborn care as advised. If we had informed the family during pregnancy that they should call us if the delivery takes place at home, we may be able to reach their home in time to be present at birth.

In the case of **institutional delivery**, we will visit the family as soon as we come to know that they have returned home from the hospital.

If we are **present at birth, we will observe**:

- Whether the attendants have washed their hands well before assisting childbirth
- Whether newborn is wiped dry immediately after birth, placed naked on the mother's abdomen or chest and covered with a clean and dry cloth.
- Whether the cord is cut with a new blade, and nothing is applied to it.
- Whether the newborn is breastfed as soon as possible after birth.
- Whether the baby is attached to mother's breast and started suckling.
- Whether it is preterm or full term-birth and whether the baby's weight is adequate. If the newborn is weak then support the family in providing extra care that the baby needs such as KMC, frequent breastfeeding, maintaining hygiene and cleanliness.
- Wherever necessary, we will intervene and help the attendants to do the right thing.

If we are **unable to be present at birth**, we must visit the family at their home to observe the following at the earliest opportunity:

- Is the baby being kept adequately warm?
- Has anything been applied to the cord?
- Has the baby been given breastfeeding already? Has the baby been fed anything other than breast milk?
- Is the baby too small (less than 2000 gms) or has she been born too soon (before 37 completed weeks)?
- Is the baby breastfeeding well?
- Wherever necessary, we will advise about adequate care and will support the family to reach the health service provider.

**Remember-Procedure :** Prepare a checklist the AWW need follow to ensure immediate newborn care after delivery and put it on the wall at AWC.

**Use-Procedure :** Ask AWW to discuss the steps she follows to another AWW.

**Find-Procedure :** Ask the other AWW to recollect the steps told by her fellow friend. While she should interrupt, give her hints and right guidance in order to rectify the mistakes done by her.

- **What is ovulation and fertilization? (Remember-Concept and Process)**

  Every month, a woman's ovary releases one egg which is called “ovulation”. If the woman is with her partner during this time then the thousands of sperms living in the semen of man will reach inside the woman's uterus. These sperms will now start searching for egg and one of the sperms will get attached to the egg. This process of sperm getting attached to the egg is called  and this is the beginning of “fertilization” creation of a new life.

- **What is menstruation? (Remember-Concept and Process)**

  If the egg and sperm do not get attached, then the egg is not fertilized and it will also not get implanted in the uterus. After waiting for 14 days of egg release, the soft lining of uterus breaks down, and “menstruation” begins. That is how we know that the woman is not pregnant.

- **What is implantation and pregnancy? (Remember-Concept and Process)**

  When the egg gets fertilized, it gets implanted to the soft lining inside uterus within a week. This is called “implantation”. If this happens, the uterus starts nurturing the pregnancy and there is no menstruation. Henceforth ovary will not release any egg during pregnancy.

  This explains that for a pregnancy to happen, release of egg is necessary. During pregnancy this release of egg from the ovary stops. After 6 weeks of delivery, the ovary again starts releasing egg. If the egg is released and if there is no pregnancy, menstruation starts after 14 days.

- **How does a woman get pregnant even before her menstruation starts after delivery? (Concept and Procedure)**

After delivery, as the uterus and ovary recover from the previous pregnancy, they become active at some point, and an egg is released. This is followed 14 days later by menstruation. The start of menstruation tells us that the egg must have been released 14 days earlier. The earliest the eggs start releasing from the ovary after a delivery, is one and a half months or six weeks.

Case 1 : The first egg after delivery was released in the eight month, and then every month. In the first three months, the eggs were not fertilized, so she menstruated 14 days after each egg release. The egg that was released the fourth time, was fertilized, so there was no menses in the fourth month. She had become pregnant sometime soon after the fourth egg release. This is what happens most commonly.

Case 2 : The first egg was released in the third month after delivery, and then one in each month thereafter. She was not staying with her husband, so the eggs did not get fertilized and she menstruated every month, 14 days after each egg released. In this case, the cycle began early after delivery.

Case 3 : The first egg was released in the third month, and was immediately fertilized by her husband. So, she never menstruated, and realized that she was pregnant only when she experienced other signs of pregnancy.

**Use and Find-Concept :** Bring three mothers with different case scenarios and ask them to discuss their scenario to the AWW. AWW explains why such things happened? What she could have done before in order to prevent it ?

- **What discussions should the AWW ask couples to start talking to each other, regarding planning a next pregnancy ? (Facts)**

  Have you thought of when do you want your next baby?
  Have you been talking to each other about this?
  Have you thought of what to do to make sure you get pregnant again only when you want to?
  Are you aware about the risks associated with unplanned pregnancy?

AWW can also tell them that it is difficult to say when will a woman start menstruating again after giving birth, and that there is always the danger that pregnancy can start even without menstruating again.

- **What should the AWW tell the parents about the benefits of planning a pregnancy? (Facts)**

  Contraception allows people to attain desired number of children and determine the spacing of pregnancies. It reduces the need for abortion, especially unsafe abortions. It can prevent closely spaced and ill-timed pregnancies and birth.

  If they show interest, we can tell them that if they can decide within six weeks after delivery, there are several ways in which the next pregnancy can be prevented or postponed.

- **What to advise if the parents are sure that they definitely do not want any more children? (Facts)**

  AWW can advise them to have the delivery in a hospital where PPIUD or PPTL services are available. PPIUD is the placement of IUD (Cu- T) immediately after delivery, and PPTL is female sterilization immediately after delivery. The family can be linked to ANM or any other health service provider for further information and advice in this regard.

# Chapter 4 Choosing a learning pedagogy

## 4.1 How to choose learning pedagogy ?

There are a number of learning theories available (Learning theories, 2018) which tells how to effectively teach, so that the learning happens effectively. The theory which considers different information type and learning objectives, for developing pedagogy was chosen for this exercise.

## 4.2 Bloom's taxonomy as a pedagogical tool

Bloom's taxonomy is a framework for hierarchical mapping, of the cognitive levels of learning. The cognitive domain of learning involves knowledge and the development of intellectual skills (Bloom, 1968). There are six categories, or levels of learning difficulties. The first should be mastered before starting the next one. They are arranged in increasing order of difficulty, from lower to higher order

- Remembering – Recognizing, listing, describing, identifying, retrieving, naming, locating, finding
- Understanding – Interpreting, Summarizing, inferring, paraphrasing, classifying, comparing, explaining, exemplifying
- Applying – Implementing, carrying out, using, executing
- Analyzing – Comparing, organizing, deconstructing, Attributing, outlining, finding, structuring, integrating
- Evaluating – Checking, hypothesizing, critiquing, Experimenting, judging, testing, Detecting, Monitoring
- Creating – designing, constructing, planning, producing, inventing, devising, making

Following this framework, one can choose the level of learning based on learning objectives. Then, an instructional queue is formed to be performed, in order to achieve the intended results.

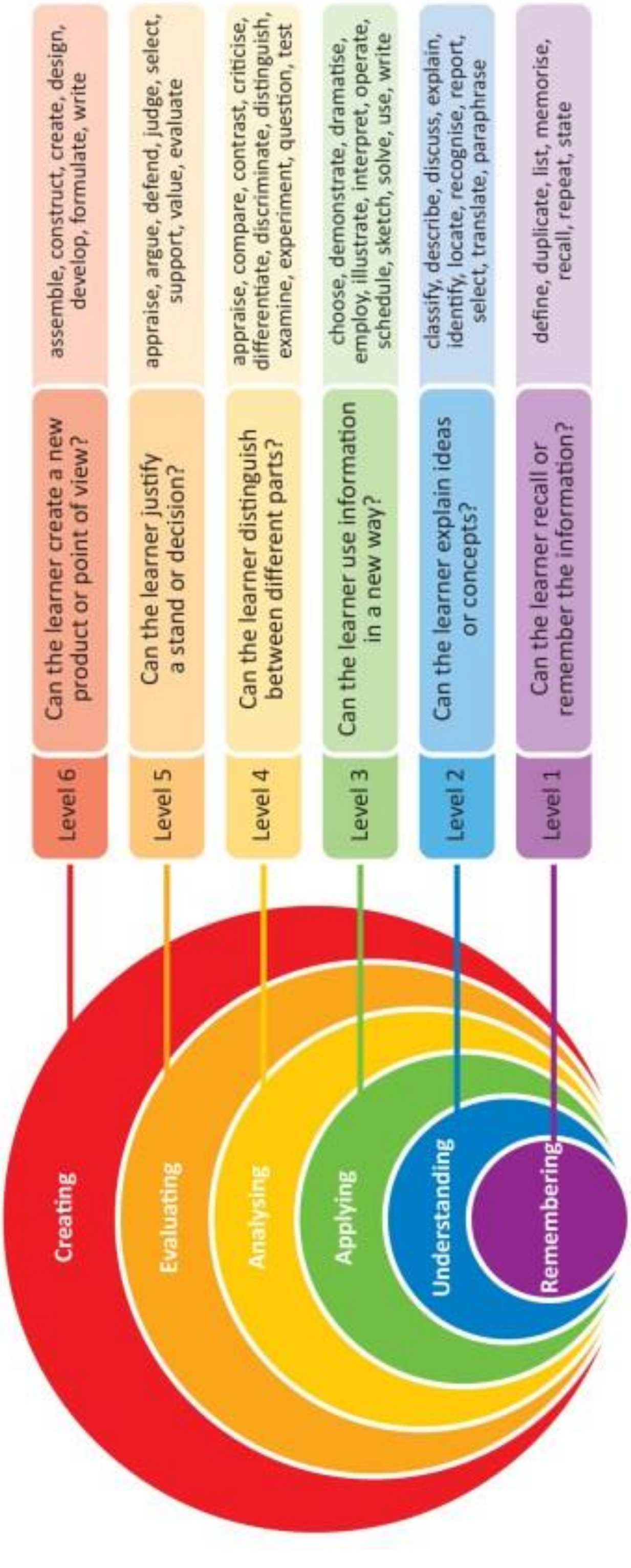

| | |
|---|---|
| **Level 6** | Do any other disease have a similar conceptual idea ? Can this concept be applied to explain other diseases ? |
| **Level 5** | Though a pregnant mother is eating enough, still her hemoglobin level is down. Explain why is it happening ? |
| **Level 4** | Consume Food > Rich in Iron > Form Hemoglobin > Embed Oxygen > Form Oxyhemoglobin > Run through Blood stream > Reaches Organs > Oxygen Released > Proper functioning of the Organs > Healthy Body |
| **Level 3** | If oxygen is insufficient, the organ suffers, muscles get tired easily, heart tries to pump the blood faster to send more oxygen, thus increasing the pulse rate |
| **Level 2** | Hb helps transport oxygen from our lungs to all organs through blood, by circulating in the body with every beat of heart |
| **Level 1** | Anemia is caused due to iron deficiency |

**Bloom's taxonomy (Understanding the learning process)**

Figure 4.1: Bloom's taxonomy (Understanding the learning process)

## 4.3 Strategy for developing engaging learning content

It is very easy to dump information on to the web in this era of World Wide Web. Many educational sites are just information dumps and one need to shovel the course out of the web. The effectiveness of instruction in most educational medium has gone down because they violate the principles that are important for learning to take place.

- Firstly, we need to show AWWs what we are trying to teach them, instead of showing instructions.
- Secondly, we need to give AWWs an opportunity to practice what they've learned. Reading and answering MCQs and assessing learning or remembering will not do. But getting them involved in classifying instances or getting involved in actually carrying out a procedure or involving them in making predictions or troubleshooting in more complex tasks might get them involved.

Motivation is perhaps the most important factor determining success in digital training. We often have a false concept of what motivation is. We think motivation comes from animation, games and edutainment, when in fact this maybe attention getting but real motivation comes from learning, when AWWs are able to do something they weren't able to do before. Real learning takes place when AWWs implement basic instructions, they recently learnt, which include demonstrating what a mother or child is going to learn having the AWW actually do what they are learning and doing this in the context of a real-world problem. AWWs take interest in trainings because they would be able to do things that they couldn't do before. If we can identify what that is and we can present that as the very first part of our instruction and say that by the end of this course or module you'll be able to do these tasks or make these predictions or perform in some way and show them what they need to do in order to complete that whole task or perform in that, their motivation is going to be greatly increased. In a summary: We need to demonstrate what's going to be learned. We need to give AWWs a chance to apply what they are going to learn. We need to do that in the context of real-world problems. Demonstration, application and problem-cantered approach needs to be taken.

# Chapter 5 Conclusion

Through this study, the contents of ILA have been structured for instructional or learning effectiveness, following Component Display Theory for content analysis and Bloom's taxonomy for a focused learning pedagogy.

It had been always kept in mind that, the learning content would be provided into some form of online instruction or gamified learning content, which the AWW can repeatedly go through and learn from it. The designer could find precisely, which part of the instruction need to be changed and the reason, why that content type is not working out for a particular AWW. Breakings down of modules, also allow AWWs to go through a topic multiple times as per need or importance. The designer could find precisely, which topics should be more focused into and which are required or demanded more for a particular context.

This analysis framework will help in designing learning course modules more efficiently and effectively. This approach can be used for designing similar instructional content from any learning material. We hope this would be used by other researchers or instructional designers to design course modules in their respective domains.